\documentclass{aa}

\usepackage{graphicx}

\bibpunct{(}{)}{;}{a}{}{,}
\usepackage{hyperref}
\hypersetup{breaklinks=true,
    citebordercolor=0.086 0.709 0.980,
    linkbordercolor=0.086 0.709 0.980}
\begin{document}
    \title{Late gas released in the young Kuiper belt could have significantly contributed to the carbon enrichment of the atmospheres of Neptune and Uranus}

   \subtitle{}

   \author{Paul Huet\inst{1}\thanks{E-mail: paul.huet@obspm.fr} \and Quentin Kral\inst{1} \and Tristan Guillot\inst{2}}

   \institute{LIRA, Observatoire de Paris, Université PSL, Sorbonne Université, Université Paris Cité, CY Cergy Paris Université, CNRS, 92190 Meudon, France\\
   \and
   Université Côte d’Azur, Laboratoire Lagrange, OCA, CNRS UMR 7293, Nice, France\\
   }

   \date{Received 21 Mars 2025, Accepted 2 June 2025.}

  \abstract
   {Exo-Kuiper belts have been observed for decades, but the recent detection of gas in some of them may change our view of the Solar System's youth. Late gas produced by the sublimation of CO (or CO$_2$) ices after the dissipation of the primordial gas could be the norm in young planetesimal belts. Hence, a gas-rich Kuiper belt could have been present in the Solar System. The high C/H ratios observed on Uranus and Neptune could be a clue to the existence of such late gas that could have been accreted ontoyoung icy giants.}
   {The aim of this paper is to estimate the carbon enrichment of the atmospheres of Uranus and Neptune caused by the accretion of the gas released from a putative gas-rich Kuiper belt.
   We want to test whether a young massive Kuiper belt such as that usually assumed by state-of-the-art models can explain the current C/H values of $\sim 50-80$ times the protosolar abundance for Uranus and Neptune.}
 {We develop a model that can follow the gas released in the Kuiper belt as well as its viscous evolution and  its capture onto planets.
 We calculate the final C/H ratio and compare it to observations. We study the influence of several important parameters such as the initial mass of the belt, the viscosity of the disc, or the accretion efficiency.}
   {We find that assuming a primordial Kuiper belt with a mass of tens of earth masses leads to significant CO gas accretion onto the giants, which can lead to high C/H ratios, especially for Uranus and Neptune.
   We find that an initial Kuiper belt of $\sim 50$ M$_\oplus$ could account entirely for the present-day C/H enrichment in the atmospheres of Uranus and Neptune. However, given the fact that S/H is also significantly enriched in the deep atmospheres of these planets but less than C/H, a more likely scenario is that these planets first accreted an envelope enriched in C/H and S/H in similar amounts, and that the sublimation of CO from the Kuiper belt led to an additional enrichment in C/H of perhaps 30 times the protosolar value in Neptune, and 20 times in Uranus. For the same model, the additional enrichment in C/H are 2 and 0.2 in Saturn and Jupiter, respectively.}
   {Our model shows that a relatively massive gas-rich Kuiper belt could have existed in the Solar System's youth, significantly enriching the atmospheres of Uranus and Neptune with carbon.
   Late gas accretion and its effect on outer giant planets metallicities could be a universal scenario, also occurring in extrasolar systems. Observations of sub-Jupiter exoplanets could provide very useful information to better constrain this scenario,
   with an enrichment in carbon and oxygen (for warm-enough planets) compared to other elements that should be inversely proportional to their envelope mass.}


   \titlerunning{The effect of the accretion of late gas on the atmospheric compositions of Neptune and Uranus.}

   \maketitle

\section{Introduction}

A considerable number of exo-Kuiper belts, also referred to as debris discs, have been observed over the past few decades.
First detected by their infrared excess due to the dust emission they produce~\citep{aumann_discovery_1984, eiroa_cold_2010}, they can now be imaged at high resolution with e.g.\ ALMA~\citep{ sepulveda_reasons_2019}.
Probably one of the most unexpected recent results was the discovery of gas (mostly CO and usually referred as late gas) in relatively old (10 - 700 Myr) exo-Kuiper belts orbiting around main-sequence stars~\citep{moor_molecular_2017, macgregor_complete_2017}.
To date, almost 30 debris discs containing gas have been observed, among which some have a considerable mass of CO~\citep[up to 0.1 M$_\oplus$,][]{moor_new_2019}.
Observations indicate that CO, C, and O gas species are present~\citep{cataldi_herschelhifi_2014, brandeker_herschel_2016} and the gas composition is expected to be dominated by those species rather than hydrogen or helium~\citep{hughes_radial_2017}.

Models indicate that this late gas is most likely not a remnant of the protoplanetary disc gas, but rather a secondary phenomenon, where gas is released from volatiles contained in planetesimals in these debris belts~\citep{kral_predictions_2017, kral_imaging_2019}.
Hence, a considerable quantity of gas is produced in the belt over an extended period of time (approximately 100 Myr), as opposed to the typical several Myr for protoplanetary discs.
It is anticipated that this gas will viscously spread both inwards and outwards, plausibly because of magnetorotational instability or molecular viscosity~\citep{kral_magnetorotational_2016, cui_dynamics_2024}.

As these systems are mature, they are expected to contain planets that are already formed, especially giants, which are potentially close to the belts. In fact, thanks to direct imaging, such planets have already been observed in debris disc systems~\citep{lagrange_constraining_2009, lagrange_evidence_2019}. It is then expected that the planets will be within reach of the viscously spreading gas disc and that some of this late gas will be accreted onto the planets. The analytical model developed by~\cite{kral_formation_2020} suggests that the accretion of this late gas onto planets is highly efficient, given that it occurs over an extended period and that the atmospheres have sufficient time to cool.

The late gas release observed in relatively young extrasolar systems possibly also happened in the young Solar System. However, today's Kuiper belt (KB) is considerably less massive than the aforementioned debris discs and gas release may be very low. Although a certain amount of CO could still be released in the current KB, the solar wind would expel this low-density gas from the belt, making it difficult to detect~\citep{kral_molecular_2021}. However, our solar system is considerably more ancient than those with extrasolar gas detected. It is reasonable to assume that the Kuiper belt would once have been much more massive and thus produce a larger quantity of gas, sufficient to overcome the solar wind and, like in some extrasolar systems, capable of spreading viscously inwards as far as Neptune and Uranus, and even Saturn and Jupiter. Indeed, the current low-mass Kuiper belt is expected to be a remnant of a much larger belt that was dispersed during the early stages of the solar system's formation by the influence of Neptune and Uranus, making it likely that it resembled the gas-rich belts observed today~\citep{bottke_collisional_2023}.

One of the most studied models of the evolution of our Solar System designed to reproduce its current architecture may be the Nice Model~\citep{tsiganis_origin_2005, morbidelli_chaotic_2005, gomes_origin_2005}. It was progressively enriched over time to solve fine-tuning issues~\citep{levison_late_2011}, timing issues~\citep{nesvorny_statistical_2012} and get a more precise representation of the primordial disc~\citep{griveaud_solar_2024}. The Nice model suggests that the system was in a more compact configuration than today, with planets within 16 au. The planets were then pushed on their current orbits after an instability between Jupiter and Saturn via resonant interactions between them. However, to get the model working and push Jupiter and Saturn into resonance, the initial belt needs to have a mass between 20 to 50 M$_\oplus$ in place of the $\sim$0.1 M$_\oplus$ of the current Kuiper belt. The primordial massive belt gets depleted when Neptune migrates outwards, which captures some Kuiper belt objects into resonances and can explain the current KB architecture. Moreover, refined collisional models such as~\cite{bottke_collisional_2023} show that the Nice model can explain the size distributions of Trojans, of craters on the moons Europa and Ganymede, as well as many features of the current Kuiper belt.

There are also alternative models that may explain the broad strides of the evolution of the Solar System. For instance,~\cite{liu_early_2022} argue that as the disc dissipates from the inside out via photoevaporation, some ``rebound'' effect may happen that could push the outer ice giants outwards and destabilize the system thus causing outward migrations for Jupiter, Saturn, Uranus, and Neptune. In the end, after the dissipation of the protoplanetary disc, the system is already in an extended configuration close to the current one, unlike the Nice model. In this case, a belt's mass down to 5 M$_\oplus$ is enough to reproduce the current planet orbits.

If the primordial Kuiper belt was indeed massive as currently assumed by most models, the gas release rate could have been similar to what is observed in some extrasolar systems, which may have important consequences on Uranus and Neptune. Indeed,
the icy giants are the closest to the Kuiper belt, and they are expected to accrete this late gas, which may significantly alter the composition of their atmospheres, as will be shown in this paper.

The metallicity of the atmospheres of Neptune and Uranus is constrained to be highly super-solar, with a C/H ratio of $\sim \, 50-80$ times the protosolar metallicity~\citep{atreya_deep_2020, guillot_giant_2023}.
To explain such an abnormally high metallicity, the current models posit that the planets formed in proximity to the CO ice line by pebble accretion~\citep{ali-dib_measured_2014, mousis_recipes_2024, mousis_insights_2024}.
The pebbles in this area are expected to be enriched in carbon. Pebbles drift inward because of their interaction with primordial gas and then sublimate when they cross the ice line. Unless otherwise stated, the term "ice" in this paper refers exclusively to CO ice.). The sublimated gas viscously spreads inwards and outwards. Then a part of it condensates onto the icy pebbles that have not yet crossed the CO iceline. When Uranus and Neptune are close to the ice line, they accrete pebbles enriched in carbon, which can enhance the C/H ratio in their atmospheres. However, even though those pebbles might pollute the Hydrogen-Helium atmospheres of Uranus and Neptune, there are still large uncertainties about their final impact~\citep{mousis_insights_2024}.

The atmospheres of Neptune and Uranus are $\sim 1$ to 4 M$_\oplus$, i.e.\ much lighter than the total masses of the planets~\citep{guillot_giant_2023}.
Accreted late CO gas will be mixed in those relatively light atmospheres, which might significantly contribute to this observed super-solar metallicity as will be explored in further detail in this paper.

After introducing our numerical model in section~\ref{section_Methode}, we study the carbon enrichment of Uranus and Neptune with our model for different plausible scenarios of the Kuiper belt evolution in Section~\ref{section_results}.
Finally, we discuss our results in Section~\ref{section_discussion} before concluding.
Briefly, we find that the primordial belts tested produced a significant amount of gas and that the most efficient solar system formation model that can enrich the atmospheres of Uranus and Neptune in carbon at the observed level is that derived from the most recent extension of the Nice model~\citep{griveaud_solar_2024}.

\section{Method} \label{section_Methode}

The aim of this paper is to investigate the accretion of gas produced in the Kuiper belt early in the solar system evolution onto the solar system's outer planets (Jupiter, Saturn, Uranus, and Neptune) over the entire lifetime of the late gas disc.
Debris discs may persist for several Gyr, like in our solar system, and the gas in it is thus expected to be present for a couple hundred million years at the minimum~\citep{matra_detection_2017, moor_molecular_2017}.

\subsection{Viscous diffusion}
\subsubsection{The viscous model}
During the large time span of the gas in the debris disc phase, it will have time to viscously evolve~\citep{kral_self-consistent_2016,kral_magnetorotational_2016}. However, given current numerical constraints, it is not realistic to run hydrodynamic simulations over such an extended period of time. In this context, we adopted the numerical methodology proposed for debris disc by, e.g. \cite{moor_new_2019,marino_population_2020}, and solve a one-dimensional viscous diffusion equation for a fluid made up of a mixture of chemical elements, which orbits the star. We assume that gas is initially released as CO. Its photodissociation caused by both the interstellar radiation field (ISRF) and stellar photons results in the partial conversion of this gas into carbon and oxygen, hence the use a multi-species model. This means that each fluid component has its own dynamic influenced by the other species.

In our model, we assume that the viscous parameter $\alpha$ \citep{lynden-bell_evolution_1974} is identical for all species, that the disc is axisymmetric and that the vertical scale height is defined, for a given radius $R$, by $H(R) = c_s/\Omega$, where $\Omega = \sqrt{\frac{GM_*}{R^3}}$ is the angular speed and  $c_s = \sqrt{\frac{k_B T}{\mu m_p}}$ is the sound speed. $M_*$ is the mass of the central star, $T$ is the gas temperature at radius $R$, $\mu$ is the mean molecular weight, $G$ and $k_B$ are respectively the gravitational and the Boltzmann constants, and $m_p$ is the proton mass.
Let us also define $\Sigma$ the total surface gas density and $\Sigma_i$ the surface density of the different species $i$ such that
\begin{equation}
    \Sigma = \sum_i \Sigma_i.
    \label{sum_sigmas}
\end{equation}

\noindent Our simulations will account for three species, namely CO, C and O.
The radial velocity of each species has two components: the global fluid radial velocity $v_r$ and the diffusive flux $v_{ri}$ which is the velocity due to the diffusion between species.

In our model, $\dot{\Sigma}_i(R, t)$ is the gas generation/destruction rate for species $i$. This term is the sum of the gas generation rate in the belt, and of its destruction rate due to photodissociation and planetary accretion, i.e. all the non-diffusive terms. The prescriptions for $\dot{\Sigma}_i(R, t)$, depending on the species, will be described more precisely in subsection~\ref{pd}.
The gas production rate $\dot{\Sigma}_{\rm CO}$ is obtained via the total gas mass production rate $\dot{M}_{\rm CO}$ calculated in subsection~\ref{warm} and~\ref{col}. We assumed a radial power law distribution for the belt of -3/2~\citep[as in e.g.][]{kral_imaging_2019}. Therefore we get $\dot{\Sigma}_{\rm CO}(r) = \frac{\dot{M}_{\rm CO}}{S_{\rm norm}} \left(\frac{r}{a_0}\right) ^{-3/2}$ with $a_0$ and $\Delta a$ the belt's semi-major axis and radial extension, respectively, and where $S_{\rm norm}=\int_{a_0-\Delta a / 2}^{a_0 + \Delta a / 2} 2 \pi \left(\frac{r}{a_0}\right)^{-3/2} r \, dr=4 \pi a_0^2 \left( \sqrt{1 + \frac{\Delta a}{2 a_0}}-\sqrt{1 - \frac{\Delta a}{2 a_0}} \right)$ is the normalisation factor.

The mass conservation of gas in the disc leads to (in cylindrical coordinates)

\begin{equation} \label{mass_cons}
    \frac{\partial \Sigma_i}{\partial t} = - \frac{1}{R} \frac{\partial}{\partial R}(R v_r \Sigma_i) - \frac{1}{R} \frac{\partial}{\partial R}(R v_{ri} \Sigma) + \dot{\Sigma}_i(R, t),
\end{equation}

\noindent while the radial velocity can be obtained from the momentum conservation since there is a friction torque defined via the viscous coefficient $\nu = \alpha c_s H$, where $\alpha$ is the viscous parameter~\citep{shakura_black_1973} assumed to be constant in our disc
 \begin{equation}
     \Sigma v_r = - \frac{3}{\sqrt{R}}\frac{\partial}{\partial R}(\nu \Sigma \sqrt{R}).
     \label{momentum_cons}
 \end{equation}

We consider $\alpha$ values ranging from $10^{-3}$ to $10^{-1}$ (see Figure~\ref{fig_simu}). Indeed, for debris discs, both turbulent viscosity, such as that caused by the magnetorotational instability, and molecular viscosity result in higher values of $\alpha$ than those observed in protoplanetary discs~\citep{kral_magnetorotational_2016, cui_dynamics_2024}.

Following~\cite{charnoz_planetesimal_2019}, the interspecies diffusion velocity equation is given by

\begin{equation}
    v_{ri} = - \nu \frac{\partial}{\partial R}\left( \frac{\Sigma_i}{\Sigma}\right).   \label{diffusive flux}
\end{equation}

 Using equations~\ref{sum_sigmas},~\ref{mass_cons},~\ref{momentum_cons} and~\ref{diffusive flux}, we can retrieve the usual mono-species viscous evolution equation for $\Sigma$ defined classically in~\cite{lynden-bell_evolution_1974} for $\dot{\Sigma}(R,t)=0$.

\subsubsection{The numerical approach}
The set of partial differential equations we need to solve has no general analytical solution, and must be solved numerically. In order to achieve this, the spatial domain is discretized on a grid with $n$ cells via the finite difference method using centered numerical derivatives. Subsequently, we obtain $n$ differential equations of order 2 in time, which we turn into $2n$ order 1 differential equations in time. To solve these equations, we use an adaptive time step RK45 solver, via the \texttt{solve\_ivp} function of the SciPy package~\citep{virtanen_scipy_2020}. After testing our setup, we chose $n=8000$ to circumvent numerical instabilities and ensure global mass conservation.

We use boundary conditions for the surface densities that are similar to those set by \citet{marino_population_2020}. For the outermost cell, we use a power-law extrapolation of the surface density based on the nearest cells. Although we must truncate the disc at an arbitrary outer radius, the disc is expected to continuously decrease in surface density rather than stop as expected at steady state~\citep{kral_imaging_2019}. We thus assume a power-law extrapolation as a good physical approximate of the outer disc.
Since the computational cost increases as the step size decreases, it is impractical to model the innermost part of the disc. Therefore, we also use a power-law extrapolation for the innermost cell. However, this extrapolation may be less reliable than for the outer part of the disc, especially at the beginning of the simulation where the gas surface density varies significantly. Thus, we also add the constraint that the flux in the innermost cell cannot be greater than the flux in the previous cell via the condition $\nu \Sigma = \text{Const}$.
We verify that our numerical model, using these boundary conditions, converges to the expected analytical solution for a constant mass production rate.

\subsection{Gas produced from warming ices in the KB}\label{warm}
There are two main processes that can release gas in the Kuiper belt. First, gas can be released via collisions as explained in more detail in subsection~\ref{col}, and second, gas can be released by sublimation of warming KB objects (KBOs) over time~\citep[as proposed in][]{kral_molecular_2021}, which we study in this subsection. For our different set-ups, we find that one process always dominates the other as we will mention more specifically later.

For KBOs in the solar system, the time scale for the sublimation of CO ices and the time scale for the CO diffusion into the porous structure of KBOs is relatively short in comparison to the thermal diffusion characteristic timescale~\citep{lellouch_tnos_2013, kral_molecular_2021}.
Therefore, KBOs will take time to reach their equilibrium temperature, while the Sun's heat penetrates slowly inside KBOs in an outside-in fashion. Since the sublimation timescale is much shorter than the thermal diffusion timescale, the model will not depend on the distance to the central star, or rather only indirectly, assuming the belt is close enough to the central star for CO ice to sublimate. Let us consider a KBO with a diameter $s$ that is initially composed of refractory materials and CO ices, with a ratio of CO ice mass to refractory mass $f_{\rm ice} = M_{\rm ice}/M_{\rm refr}$ with typical values of 0.1~\citep[e.g.,][]{mumma_chemical_2011}.

At a given time $t$, the ice within the radius $r = \sqrt{Kt}$ (where $K$ is the thermal conductivity of the KBO) will remain intact but it will turn into gas above $r$. We assume a thermal conductivity $K$ of $\sim 10^{-10} \ \mathrm{m^2} \, \mathrm{s^{-1}}$ as given by observations of these cold objects~\citep{prialnik_modeling_2004, lellouch_tnos_2013}.
Because all the CO above the radius $r$ sublimates, we can calculate that during a timescale ${\rm d}t$, all of CO in the layer between $s / 2 - \sqrt{Kt}$ and $s / 2 - \sqrt{K(t + dt)}$ sublimates. Hence, the mass loss rate for a KBO of diameter $s$ as a function of time can be easily estimated~\citep[e.g.,][]{kral_molecular_2021}

\begin{equation}
\begin{split}
    \frac{{\rm d}M_{\rm CO}}{{\rm d}t}(s) &= 2 \pi \rho_{\rm refr} f_{ice} K^{3/2} \sqrt{t} \ \mathrm{if} \ t < \frac{s^2}{4 K},\\
                          &= 0 \ \mathrm{if} \ t \ge \frac{s^2}{4 K}, \\
\end{split}
\end{equation}

\noindent where $\rho_{\rm refr}$ is the density of the refractory material.

As explored in the next subsection, the gas production rate from collisions is usually much higher when the belt is young and massive. Hence, we expect the ice sublimation to only dominate when the disc becomes less massive, which happens after the massive primordial belt depletion (see subsection~\ref{subsection_belt_model}).
We assume that the KBO number density will not evolve significantly during this later phase and we can calculate the total sublimation rate at a time $t$ coming from all KBOs in the belt as

\begin{equation}
    \frac{{\rm d}M_{\rm CO}}{{\rm d}t} = \int_{{\rm max}(s_{\rm min}, \sqrt{Kt})}^{s_{\rm max}}  2 \pi \rho_{\rm refr} \, f \, K^{3/2} \, \sqrt{t} \, n(s) \, {\rm d}s,
\end{equation}
where $n(s)$ is the belt's size distribution, which will be one of the parameters we explore in this study. $\dot{M}_{\rm CO}$ can then be fed in our model via the $\dot{\Sigma}_{\rm CO}(R, t)$ variable.

\subsection{Gas produced from collisions}\label{col}

In younger, heavier belts, such as the primordial Kuiper belt we consider here, the effect of collisions onto gas production might not be negligible~\citep{thebault_collisional_2007, bonsor_secondary_2023}.
The standard model to predict gas production rates in collisionally evolving debris discs is described in \citet{kral_predictions_2017}. It assumes that the gas production rate is proportional to the solid mass loss rate. The idea is that collisions produce smaller debris and expose new surfaces that can sublimate and release gas.
We use this approach to compute the gas release rate due to collisions coupled to an analytic model for the estimation of the solid mass loss rate, able to account for several size distribution slopes~\citep{lohne_long-term_2008}. The latter model can be adapted to compute the solid mass loss rate in the primordial Kuiper belt using primordial slopes for the young Kuiper belt size distribution from \citet{bottke_collisional_2023}.
We use formula 38 of \citet{lohne_long-term_2008} to estimate the belt's mass as a function of time and then derive it numerically to obtain the solid mass loss rate $\dot{M}_{\rm belt}$. Similar to \citet{kral_predictions_2017}, we then estimate the gas generation rate to be $\dot{M}_{\rm CO}=-f_{\rm ice} \, \dot{M}_{\rm belt}$, where we assume that all ice contained in KBOs will eventually sublimate as the planetesimals are turned into dust. This is a reasonable assumption since the area of exposed KBOs increases significantly during collisions (since it produces smaller chunks with higher cross sections given assumed size distributions) and the sublimation proceeds at the surface, not deep inside the KBOs.

\subsection{Planet's influence}
In order to model three-dimensional hydrothermodynamic effects, such as planetary accretion, within our one-dimensional viscous diffusion code, it is necessary to parameterize the problem.
To do so, let us define the accretion efficiency
\begin{equation}
    f_{\rm accr} = \frac{\dot{M}_{\rm accr}}{\phi_{in}},
    \label{equation_f_accr}
\end{equation}
\noindent where $\dot{M}_{\rm accr}$ is the planet's mass accretion rate, and $\phi_{in}$ denotes the incoming mass flux.
It should be noted that the incoming mass flux is not necessarily the total mass flux at the planet's semi-major axis. Indeed, the gas disc has a considerable vertical scale height, whereas the planet has a gravitational sphere of influence of radius $R_{accr}$, beyond which the gas is not accreted. Accounting for that, the incoming mass flux to a planet of semi-major axis $a$ can be expressed using the minimum function as follows~\citep[][]{kral_formation_2020}

\begin{equation}\label{phiin}
\phi_{in}(t)={\rm min}(1, \frac{R_{\rm accr}}{H}) \  \Sigma(a,t) \ v_r(a,t).
\end{equation}

We estimate the planet's accretion radius, $R_{\rm accr}$ such that  $R_{\rm accr}={\rm min}(R_{\rm Hill}, R_B)$,
where $R_{\rm Hill} = a \ \left( \frac{M_{\rm planet}}{3M_*} \right)^{1/3}$ is the Hill radius beyond which the central star's gravity will be stronger than the planet's gravity, and $R_B = \frac{GM_{planet}}{v_r^2(a, t)}$ the Bondi radius, which is the planet's escape radius for a particle moving at radial velocity $v_r$. Note that we do not use the gas sound speed in the disc as the characteristic speed, since the planet's atmosphere is expected to be decoupled from the disc in low-density environments similar to the late stages of giant planet formation in protoplanetary discs~\citep{mordasini_characterization_2012}. We note that in our upcoming simulations, $R_{\rm Hill}$ is always smaller than $R_B$.

The gas discs under consideration have typically a lower density than protoplanetary discs and in this case,~\cite{kral_formation_2020} predict that the accretion efficiency is very high and may be close to 100\% (as will be discussed further in section \ref{section_discussion}).
Consequently, the main limiting factor of the gas accretion rate onto a planet would be determined by hydrodynamic processes in addition to the radius of influence $R_{\rm accr}$.
As demonstrated by previous studies in protoplanetary disc environments, it is challenging to accurately determine the amount of gas flowing around the planet that gets accreted~\citep{lubow_gas_2006, mordasini_characterization_2012, ormel_hydrodynamics_2015}. In addition, numerical hydrodynamic simulations have not yet been run in the debris disc phase and more refined effects may have to be accounted for.
Based on previous numerical estimates~\citep{lubow_gas_2006, mordasini_characterization_2012, tanigawa_distribution_2012, ormel_hydrodynamics_2015, lambrechts_quasi-static_2019}, we assume an accretion efficiency due to hydrodynamics processes of $f_{\rm accr}=0.5$. However, to account for uncertainties in this parameter, based on those hydrodynamic studies, we also test two alternative values of 0.1 and 0.8.

Numerically, we use sink cells to account for gas accretion onto planets. Therefore, at each time step, we compute the amount of gas mass that should be accreted onto the different planets and remove it from the gas phase similar to simulations of \citet{kral_impact-free_2024} for water released in the young asteroid belt.

\subsection{Photodissociation}\label{pd}
Our model follows the evolution of carbon monoxide outgassed in the belt as well as its photodissociation products C and O. We assume that photodissociation is mainly produced by the interstellar radiation field (ISRF).
This is because the Kuiper belt is located at a considerable distance from the Sun and that the disc's radial extent, due to radial spreading, is significantly larger than its vertical scale height. Consequently, the majority of the disc is optically thick in the radial direction and thus mostly opaque to solar radiation in the UV~\citep[e.g.,][]{kral_self-consistent_2016}.

The theoretical lifetime of CO, $t_0$, in the case of impinging photons from the ISRF is estimated to be 120 years~\citep{visser_photodissociation_2009}. However, CO can be shielded both by itself~\citep{visser_photodissociation_2009} and by neutral carbon~\citep{kral_imaging_2019}.
We interpolate the CO self-shielding function $\Theta(\Sigma_{\rm CO})$ used in our model (when no self-shielding is present, it tends to 1) from~\cite{visser_photodissociation_2009}. For neutral carbon, we estimate a critical surface density $\Sigma_{\rm crit}$ above which the optical depth to incoming vertical radiation would be superior to one.
Following~\cite{kral_imaging_2019,marino_population_2020}, we expect carbon shielding to have an exponential dependence on neutral carbon surface density, with a critical density $\Sigma_{\rm crit} \sim 10^{-7} \,\mathrm{M_\oplus \, au^{-2}}$.
In this model, we assume that the carbon ionization fraction is small, similar to~\cite{marino_population_2020}. This assumption is less reliable for the discs with low masses~\citep{kral_predictions_2017}. However, even in the extreme case where all carbon is ionized, because the gas disc is of low mass, shielding will not be important and the ionised carbon gas disc will behave exactly as the neutral carbon since it has the same molecular mass.
We find that the interspecies diffusion is negligible in our simulations leading to no differences between neutral and ionised carbon diffusion.

Finally, the CO photodissociation timescale in the disc can be estimated as
\begin{equation}
    t_{\rm ph} = t_0 \ \frac{\exp{\left[\frac{\Sigma_{\rm C}}{\Sigma_{\rm crit}}\right]}}{\Theta(\Sigma_{\rm CO})},
\end{equation} where $\Sigma_{\rm C}$ and $\Sigma_{\rm CO}$ are respectively the surface density of neutral carbon and that of CO. The photodissociation contribution to $\dot{\Sigma}_{\rm i}$ is then given for the different species by
\begin{equation}
    \dot{\Sigma}_{\rm CO} = -\frac{\Sigma_{\rm CO}}{t_{\rm ph}}
\end{equation}
\begin{equation}
    \dot{\Sigma}_{\rm C} = \frac{\mu_{\rm C}}{\mu_{\rm CO}} \ \frac{\Sigma_{\rm CO}}{t_{\rm ph}}
\end{equation}
\begin{equation}
    \dot{\Sigma}_{\rm O} = \frac{\mu_{\rm O}}{\mu_{\rm CO}} \ \frac{\Sigma_{\rm CO}}{t_{\rm ph}}
\end{equation}
where $\mu_{\rm CO}$, $\mu_{\rm C}$ and $\mu_{\rm O}$ are the molecular weight of CO, C and O, respectively.

\subsection{The Kuiper belt modelling and evolution} \label{subsection_belt_model}

\begin{figure}[htbp]
\begin{centering}
\includegraphics[scale=0.42]{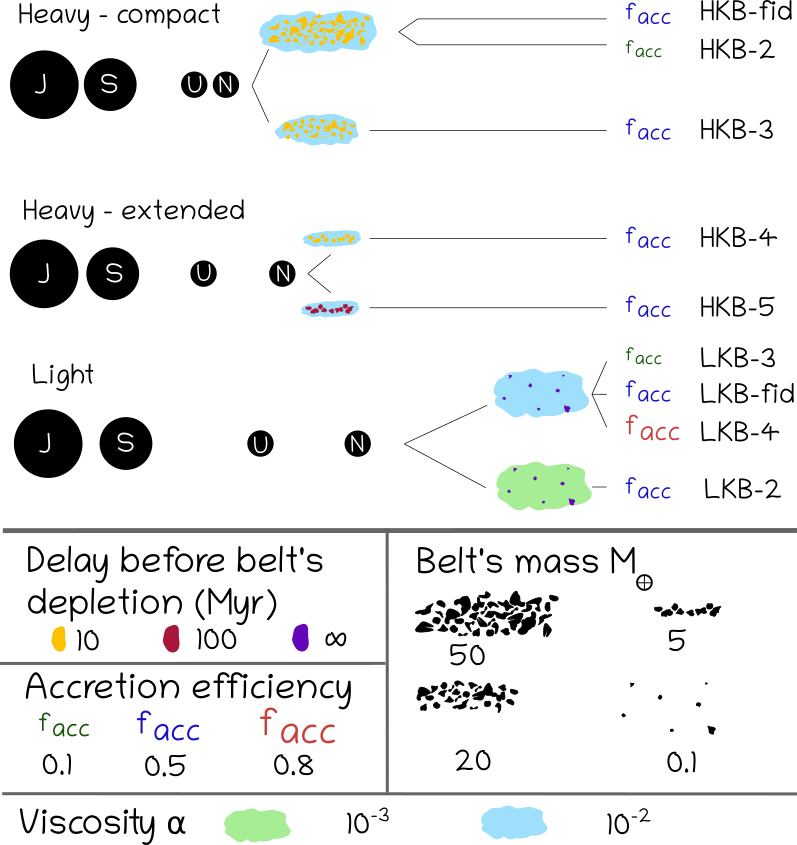}
\end{centering}
\caption{Schematic presenting a description of our set of 9 simulations. The heavy belt cases are at the top (compact configuration) and middle (extended configuration), and the light cases are at the bottom. J, S, U, and N are Jupiter, Saturn, Uranus, and Neptune, respectively. The different simulations differ in terms of the locations of planets, accretion efficiency, the viscosity of the gas, and the belt's mass. For the heavy belt configurations, the time when depletion starts is also a parameter. Each parameter has its own symbol and colour as can be seen in the bottom part of the cartoon.}
\label{fig_simu}
\end{figure}

In our simulations, we consider three different configurations of the young Kuiper belt, each corresponding to a distinct scenario for the belt's early history (see Fig. \ref{fig_simu}). The first set of simulations (HKB-fid, 2 and 3 simulations, see Tab.~\ref{table:simulations_HKB}) takes as a basis the Nice model, i.e. we start with a heavy primordial belt and planets closer to the Sun than their current positions. The second set (HKB-4, 5) takes its initial set-up from the ``Rebound'' scenario of \citet{liu_early_2022}, which also assumes a relatively massive initial belt but with planets closer to their current locations than in the Nice model. Last, we consider a set of simulations (LKB-fid, 2, 3, 4, see Tab.~\ref{table:1}) assuming that the KB was born ``light'', with a configuration similar to the current KB and planets at their current locations. This light-KB scenario can be considered as a lower limit in terms of gas production within the belt and its effect on Uranus and Neptune atmospheres.

\subsubsection{The Nice model scenario}\label{nice}

In the Nice model scenario, the primordial belt is massive ($\gtrsim 10$ M$_\oplus$) and evolves over a time $t_{\rm ev}$ of 10-100 million years before being depleted by Neptune's outer migration. Recent results suggest that this value is closer to 10 Myr, which is what we assume in our model~\citep{bottke_collisional_2023}. We thus consider that the mass depletion of the belt starts after a time $t_{\rm ev}$ and is modelled by an exponential decay with an e-folding time $t_{\rm fold}$. For the Nice model, we expect  $t_{\rm fold} \sim \, 10$ Myr~\citep{bottke_collisional_2023}, which is the value we take. We assume that the mass tends to zero at the end of the simulations, which is unrealistic but does not affect the final results, dominated by the early phase.

The planets and belt positions are also similar to those assumed in the most recent iterations of the Nice model~\citep[e.g.,][and references therein]{griveaud_solar_2024}. The primordial belt is taken to be located at 27 au from the Sun and to be 6 au wide~\citep{bottke_collisional_2023}. Neptune, Uranus, Saturn, and Jupiter are in a compact configuration, at 16, 14, 8, and 5 au, respectively. Because of gravitational interactions with the primordial KB, the outer planets move outwards and Jupiter and Saturn cross their 2:1 resonance, which excites their eccentricities and triggers an instability, which eventually depletes the KB. We do not model the migration of planets before the instability and subsequent depletion but keep the belt and planet positions fixed throughout. This is because the positions of planets are not an important parameter and do not alter our final results as discussed later. However, because of the migration of planets, planetesimals in the young belt become excited dynamically. To account for that, we assume a mean eccentricity of 0.25 in our collisional model, slightly higher than the 0.05-0.1 typically assumed~\citep[e.g.,][]{thebault_collisional_2007}.

The size distribution of the primordial KB is assumed to be different from that of the current KB. We use state-of-the-art models of the early KB for our simulations, which assume that the size distribution can be separated into three regimes: for bodies between $10^{-4}$ m and 100 km in diameter, the size distribution has a slope of -2.1, between 100 and 300 km the slope is -6, and for bodies up to 4000 km the slope is -3.5~\citep{bottke_collisional_2023}. We assume that the spatial distribution scales as $R^{-3/2}$, independent of time~\cite[similar to][]{bottke_collisional_2023}. The HKB-fid, 2 and 3 simulations only differ in the assumed accretion efficiency of gas onto planets (0.5 and 0.1) and the initial belt mass (20 and 50 M$_\oplus$ are tested). In these simulations, the gas production mechanism is dominated by collisions (see Sec.~\ref{col}).

\subsubsection{Alternative ``heavy initial belt'' scenarios}

We run another type of simulations to account for scenarios in which the primordial belt starts heavy but the planets, especially Neptune, start farther away, i.e. closer to their current positions. In this case, models do not need as heavy a belt as in the Nice model, and an initial belt under 10 M$_\oplus$ can reproduce the Solar System evolution~\citep[e.g.,][]{liu_early_2022}. We assume a typical belt mass of 5 M$_\oplus$ for those scenarios.

The ``Rebound'' scenario proposed recently in \citet{liu_early_2022} for forming the Solar System can be modelled via those simulations. In this scenario, the protoplanetary disc is photoevaporated from the inside out from a certain radius, creating a gap in the young protoplanetary disc. When the outer edge of the gap reaches Saturn, it pushes the giant outwards, which triggers an orbital compression of planets and finally an instability~\citep{liu_early_2022}. In this scenario, Neptune migrates over a smaller distance than in the Nice model. In our simulations taking as a basis this scenario, which shall be referred to as the ``extended'' configuration, Neptune's location is 23 au (instead of 16 au in the Nice model runs). For all the other 3 planets, we consider the same locations (5, 8, and 14 au) as in the compact Nice-model simulations of section~\ref{nice}. As for the compact simulations, the migration of the planets is not dynamically modelled and the planets remain at fixed positions throughout the runs, but we will show that the planets' locations only weakly affect their level of carbon enrichment. As for the eccentricity, belt position and width, planetesimal size distributions and surface density, they are taken identical as in the compact Nice-model simulations.

We run 2 simulations for this extended configuration, HKB-4 and HKB-5, which differ in their $t_{\rm ev}$ and $t_{\rm fold}$ values. The HKB-4 simulation uses the same values as that of the Nice model (10 Myr for both parameters) but for the HKB-5 simulation, we test values of 100 Myr for both $t_{\rm ev}$ and $t_{\rm fold}$. This is because the survival timescale of the primordial KB in those scenarios is not very well constrained and needs to be explored. In these simulations, the gas production mechanism is also dominated by collisions (see Sec.~\ref{col}).

\subsubsection{Light early belt scenarios}

Some less common scenarios for the formation of the solar system assume that the KB was born light with a mass and size distribution very similar to the belt we observe today~\citep{shannon_forming_2016}. We run 4 simulations (LKB-fid, 2, 3, and 4) based on this scenario, for which we assume an initial belt mass of 0.1 M$_\oplus$ and a size distribution corresponding to the current one. Using \citet{morbidelli_re-assessment_2021} we take a slope of -4 for bodies with radii between $10^{-4}$ and 30 m, of -3 for bodies smaller than 100 km and, -8 for bodies smaller than 4000 km.
For these light belt simulations, we assume the KB's surface density follows a smooth Gaussian distribution centered at $R_0 = 44 \, \mathrm{au}$ with a total width of $8 \, \mathrm{au}$. Similar to the heavy primordial belt scenarios, we do not model any dynamical effect so that the locations of the planets remain fixed throughout our simulations and are assigned to their current values of 30, 19.2, 9.5, and 5.2 au for Neptune, Uranus, Saturn, and Jupiter, respectively. We do not introduce a depletion term for the belt, whose mass only decreases because of the CO ice it loses (see Sect.~\ref{warm}).

The LKB-fid, 2, 3, and 4 simulations differ in their respective values of $\alpha$ ($10^{-3}$, $10^{-2}$), and accretion efficiency (0.1, 0.5, 0.8) as presented in Tab.~\ref{table:1}. In these simulations, the gas production mechanism is dominated by the heating of the ice in the planetesimals rather than by collisions (see Sec.~\ref{warm}). We note that to be realistic, we could take the start of those simulations as the end state of the heavy-belt scenarios after the belt depletion. However, our final results would not be affected (see discussion) and we decided to optimize CPU time by not coupling those two simulations and only use the LKB models for the scenario where the KB was born light.

\subsection{Estimation of the planet atmospheric $\frac{C}{H}$ ratios in our model}

Observations give us estimates of the atmospheric $\frac{C}{H}$ ratios for Uranus, Neptune, Jupiter and Saturn. However, our numerical model provides estimates of the accreted carbon mass and we will now describe how we turn it into a modelled $\frac{C}{H}$ ratio prediction. We consider that carbon mix into the atmospheres of the giants, which consist of a mixture of hydrogen and helium~\citep{guillot_giant_2023}. From the accreted carbon mass $M_{\rm accr}^C$ and the atmospheric mass $M_{\rm atm}$, we can then calculate the $\frac{C}{H}$ abundance ratios such that

\begin{equation}
   \frac{C}{H} = \frac{\mu_{H}}{\mu_C} \frac{M_{\rm accr}^C}{M_{\rm atm}^H}
\end{equation}

\noindent where $\mu_{H}=m_p$ and $\mu_{C}=12 \ m_p$, with $m_p$ the proton mass, and the atmospheric hydrogen mass $M_{\rm atm}^H = f_H M_{\rm atm}$, with $f_H$ the hydrogen mass fraction in the atmosphere. We assumed $f_H \sim 0.746$~\citep{asplund_chemical_2021}.

The atmospheres of Neptune and Uranus are light compared to the total masses of these planets. We assumed atmospheric masses of 1.25 -- 3.5 M$_\oplus$ and 1.6 -- 4.15 M$_\oplus$ for Uranus and Neptune, respectively~\citep{guillot_giant_2023}. We use the error bars on those atmospheric masses to calculate the error bars of the modelled $\frac{C}{H}$ ratios.

For giant planets, the total atmospheric masses are much higher. We expect the late gas to be accreted early in the life of the giant planets (before 1 Gyr), i.e. before the hydrogen-helium phase separation that only happens once the planets have cooled down~\citep{howard_evolution_2024}. It is therefore not expected that the current H/He transitions at depths corresponding to atmospheric masses of $\sim 30 M_\oplus$ for both Jupiter and Saturn~\citep{markham_stable_2024, guillot_giant_2023} could have trapped carbon in this convective upper layer. One should rather consider the total atmospheric mass. The only uncertainty then comes from the extent of the diluted solid core. We consider a typical 60 M$_\oplus$ diluted solid core for both Jupiter and Saturn~\citep{howard_jupiters_2023, mankovich_diffuse_2021, guillot_giant_2023}. Some static models of Jupiter favor a higher dilute mass of $\sim$ 50 $\%$ of the total mass~\citep{militzer_juno_2022}, leading to a larger but still small enrichment in Jupiter, given its massive envelope. We also use lower limits of 20 and 30 M$_\oplus$ for the solid core masses of Saturn and Jupiter, respectively~\citep{howard_jupiters_2023, mankovich_diffuse_2021, miguel_jupiters_2022}, which we use to derive upper limits of the total atmospheric mass carbon can mix with.

Since the atmospheric carbon enrichment in the planets is usually given compared to the protosolar abundance in the literature, we will use the notation $\left[\frac{C}{H}\right]$ when relative to protosolar abundance such that $\left[\frac{C}{H}\right] \equiv \frac{\frac{C}{H}}{\left.\frac{C}{H}\right|_{\rm proto}}$, where $\left.\frac{C}{H}\right|_{\rm proto}$ is the protosun's $\frac{C}{H}$ ratio equal to $(3.33 \pm 0.31) \times 10^{-4}$~\citep{guillot_giant_2023}.

\section{Results} \label{section_results}

Here, we explore how a primordial KB may have enhanced the metallicity of the giants of our solar system. In particular, the metallicities of Uranus and Neptune are highly super-solar with $\left[\frac{C}{H}\right]$ values around 50-80 times the protosolar abundance. However, some part of this enriched metallicity may come from enrichment during planet formation in the protoplanetary disc, due to pebble or planetesimal accretion phases~\citep{guillot_composition_2006,mousis_insights_2024}. Therefore, to give an idea of the extra metallicity needed to fully explain the current C/H values of Uranus and Neptune, we start this result section by computing an estimate of the potential early enrichment (i.e. not due to late gas accretion) using the S/H ratios in Uranus and Neptune atmospheres as a good tracer of this early enrichment. Indeed, S is not expected to be released as gas (unlike CO) and it may rather be accreted together with pebbles or planetesimals when forming the planets. The S/H ratio and the corresponding C/H coming from early accretion are therefore not expected to vary after the formation of the planets, contrary to the global C/H ratio that could increase due to late gas accretion. We can then deduce the amount of late gas needed to explain observations while accounting for some potential early formation enrichment (see subsection~\ref{SsurHsec}). In this section, we will also try the same calculation based on the D/H rather than the S/H to explore if we can extract more information about this early enrichment.
In our simulations, we explore whether we can explain the extra potential enrichment as coming from gas released in the young primordial Kuiper belt and then spreading towards the giants with some of it captured by the different planets. Because the S/H and D/H values are not necessarily reliable tracers of early accretion (as error bars are quite large, see later), we also assess whether the observed $\left[\frac{C}{H}\right]$ values can be explained from late gas without considering any primordial enrichment. As a matter of fact, we will see that the answer is positive.

We estimate the gas production rate originating from slowly warming young KBOs above the sublimation temperature of CO to be significantly lower than that originating from collisions. Hence, to get a lower estimate of the atmospheric carbon enrichment expected for Uranus and Neptune (as well as for Jupiter and Saturn), we will continue this result section by first showing the results for the light Kuiper belt scenario (subsection~\ref{subsection_light_belt}), i.e. a belt very similar to the current KB. We will finish this section with subsection~\ref{HKBpart}, which is the heart of this paper showing the main results concerning the plausible important enrichment of Uranus and Neptune by late gas when assuming a massive primordial KB.

\subsection{Estimation of the C enrichment coming from early planetary formation in the protoplanetary disc}\label{SsurHsec}

We start this results section by estimating the carbon enrichment from early accretion of planetesimals or pebbles into the protoplanetary disc using two different observables: a) the S/H ratio, and b) the D/H ratio. It is necessary to do this before presenting the results of our simulations because we want to know how much carbon enrichment is required from late gas compared with the amount that could have been supplied earlier during planet formation.
The following estimations rely on the assumptions that during late gas accretion, only CO and its photodissociation products (C and O) are accreted, whereas many other elements are accreted with carbon during the planets' formation. During planet formation, solids pollute the atmosphere with heavy elements such as carbon and sulfur, as well as deuterium. The late gas accreted later is produced by the sublimation of CO ices and we do not consider any water sublimation. We do not expect any dust accretion since Neptune is outside the dust disc, and the drag between dust and gas is small in debris discs because the mass of the secondary gas disc is much smaller than that of a protoplanetary disc. Therefore, we do not expect dust to be able to diffuse inwards with gas up to the planet locations~\citep{takeuchi_dust_2001, olofsson_halo_2022}.

\subsubsection{Using the S/H ratio to estimate carbon enrichment from planetary formation}\label{soverh}

In our model, the accreted material consists only of carbon and oxygen, since gas is produced by sublimation of CO ices. However, during their formation in the protoplanetary phase, Uranus and Neptune were also enriched in carbon as well as other heavy elements such as phosphorus (P) or sulphur (S) because of infalling planetesimals or pebbles~\citep[e.g.][]{guillot_composition_2006,mousis_insights_2024}. This is a key difference in being able to determine the amount of carbon that may have come from early planetary formation rather than late gas. Therefore, we can use the amount of S observed in Uranus and Neptune and extract the corresponding amount of carbon associated with S at that early time. By comparing this with the current carbon enrichment, we can then access the amount of extra carbon needed beyond the protoplanetary disc phase to explain the observations, which is the value that our simulations should try to reproduce rather than the currently observed C/H ratio. This is the essence of the equations presented below.

Estimates of the S quantity are based on measurements of H$_2$S molecules detected at radio wavelengths deep in the atmospheres of these planets~\citep{molter_tropospheric_2021, tollefson_neptunes_2021}. Estimates of the carbon quantity are based on CH$_4$ abundances at $\sim$ 1 bar~\citep{sromovsky_methane_2019,karkoschka_haze_2011}. CH$_4$, H$_2$S, and NH$_3$ are the only major equilibrium species detected in Uranus and Neptune. The NH$_3$ abundance is very low and probably not a reliable guide because it is removed by meteorological processes linked to water condensation~\citep[``mushballs'', see][]{guillot_storms_2020}.

For Uranus, observations lead to a $\left[\frac{C}{H}\right]$ ratio between 44 and 74, and a $\left[\frac{S}{H}\right]$ ratio between 29 and 47. For Neptune, the $\left[\frac{C}{H}\right]$ ratio is measured to be between 55 and 92, and the $\left[\frac{S}{H}\right]$ ratio between 41 and 68 considering the case of wet adiabats~\citep{tollefson_neptunes_2021, guillot_giant_2023}. The $\left[\frac{S}{H}\right]$ ratios are smaller than the $\left[\frac{C}{H}\right]$ ratios or wet adiabats, which could be a sign of late gas accretion onto Uranus and Neptune. This is not the case for dry adiabats, but since this model is less favoured in the analysis of~\cite{tollefson_neptunes_2021}, we only use the case of wet adiabats to estimate $\left[\frac{C}{H}\right]$ and $\left[\frac{S}{H}\right]$ in this paper. From these values, an estimate of carbon enrichment due to the early formation can be obtained and we can then work out an estimate of the atmospheric carbon enrichment needed from late gas when comparing to observed C/H ratios, as we now show.

Let us take an arbitrary volume $V$ of atmosphere. This volume contains a number $n_s$ of sulphur atoms, $n_C$ of carbon and $n_H$ of hydrogen. We assume that all the hydrogen atoms come from the protosolar nebula gas, neglecting the hydrogen enrichment due to the accretion of solid materials in the protoplanetary phase. Sulphur is assumed to come from the protosolar nebula and from early accreted materials during the planetary formation in the protoplanetary disc such that $n_S^{atm} = n_S^{proto} + n_S^{ppd}$.
For carbon, there is an extra term from the accretion of late gas produced in the KB such that $n_C^{atm} = n_C^{proto} + n_C^{ppd} + n_C^{KB}$. Since we assume the number of hydrogen atoms to be constant, we obtain $\left. \frac{S}{H} \right|_{atm} = \frac{n_S^{atm}}{n_H}$ and $\left. \frac{C}{H} \right|_{atm} = \frac{n_C^{atm}}{n_H}$ ratios or equivalently $\left. \frac{S}{H} \right|_{atm}= \left. \frac{S}{H} \right|_{proto} +\left. \frac{S}{H} \right|_{ppd}$, $\left. \frac{C}{H} \right|_{atm} = \left. \frac{C}{H} \right|_{proto} + \left. \frac{C}{H} \right|_{ppd} + \left. \frac{C}{H} \right|_{KB}$.
Finally, when rewriting those terms relatively to the protosolar $\frac{S}{H}$ and $\frac{C}{H}$ ratios, we find

\begin{equation}
    \frac{\left.\frac{S}{H} \right|_{atm}}{\left.\frac{S}{H} \right|_{\rm proto}} =\frac{\left.\frac{S}{H} \right|_{\rm ppd} + \left.\frac{S}{H} \right|_{\rm proto}}{\left.\frac{S}{H} \right|_{\rm proto}}.
\end{equation}
\begin{equation}
    \frac{\left.\frac{C}{H} \right|_{atm}}{\left.\frac{C}{H} \right|_{\rm proto}} = \frac{\left.\frac{C}{H} \right|_{\rm ppd} + \left.\frac{C}{H} \right|_{\rm KB}+ \left.\frac{C}{H} \right|_{\rm proto}}{\left.\frac{C}{H} \right|_{\rm proto}}.
\end{equation}

To get an estimate of $\left.\frac{C}{H} \right|_{\rm KB}$ (which is our final goal), we use the values of $\left.\frac{S}{C} \right|_{\rm ppd}$, the $\frac{S}{C}$ ratio of accreted material during planet formation in the protoplanetary disc, and of $\left.\frac{S}{C} \right|_{\rm proto}$, the protosolar $\frac{S}{C}$ ratio. We find

\begin{equation}\label{eqCoH}
    \frac{\left.\frac{C}{H} \right|_{\rm KB}}{\left.\frac{C}{H} \right|_{\rm proto}} =\frac{\left.\frac{C}{H} \right|_{atm}}{\left.\frac{C}{H} \right|_{\rm proto}} - \frac{\left.\frac{S}{C} \right|_{\rm proto}}{\left.\frac{S}{C} \right|_{\rm ppd}} \left( \frac{\left.\frac{S}{H} \right|_{atm}}{\left.\frac{S}{H} \right|_{\rm proto}} - 1\right) -1.
\end{equation}

While $\left.\frac{S}{C} \right|_{\rm proto}$ is measured to be $4.6 \times 10^{-2}$~\citep{guillot_giant_2023}, we do not have precise estimates of $\left.\frac{S}{C} \right|_{\rm ppd}$. Based on models of planet formation we can estimate that its value may be close to that of the protosolar abundance~\citep[e.g.,][]{mousis_insights_2024}. Therefore, we assume $\frac{\left.\frac{S}{C} \right|_{\rm proto}}{\left.\frac{S}{C} \right|_{\rm ppd}} \sim 1$ to compute our predictions on early enrichment.

\begin{figure}[tbh]
\begin{centering}
\includegraphics[scale=0.57]{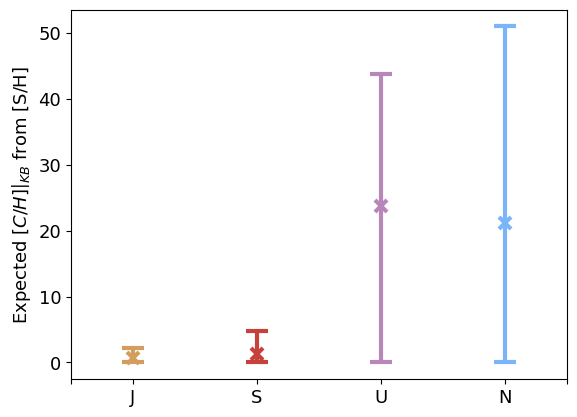}
\end{centering}

\caption{\label{fig:ch_pred}Estimations of the contribution of late gas to the atmospheric [C/H] ratios for Jupiter (J), Saturn (S), Uranus (U), and Neptune (N) calculated from the observed [C/H] and [S/H] (see subsection \ref{soverh}). The crosses show the mean values and the bars show the uncertainties, noting that the minimal values go to zero.}

\end{figure}

Finally, using Eq.~\ref{eqCoH}, we find that $\left[\frac{C}{H}\right]_{\rm KB}^{U} = 23.68_{-26.74}^{20.04}$ for Uranus, and $\left[\frac{C}{H}\right]_{\rm KB}^{N} = 21.13_{-33.99}^{29.84}$ for Neptune. For Jupiter and Saturn, we obtain $\left[\frac{C}{H}\right]_{\rm KB}^{J} = 0.64 \pm 1.54$, and $\left[\frac{C}{H}\right]_{\rm KB}^{S} = 1.33 \pm 3.41$, respectively. The results are summarized in Figure~\ref{fig:ch_pred}.

\subsubsection{Using the D/H ratio to estimate carbon enrichment from planetary formation}

The D/H ratios of Uranus and Neptune might also be used to get another estimation of the gas-rich Kuiper belt contribution to the atmospheric carbon enrichment for Uranus and Neptune. In fact, the D/H values for Uranus and Neptune are intermediate between the protosolar and cometary values and the difference with the protosolar abundances may come from the first materials accreted (planetesimals, pebbles) in the atmosphere of these planets during the protoplanetary phase. The calculation presented here below is very similar to the S/H ratio described above. In this case, the late gas only contributes carbon and oxygen, while the early planetary formation in the protoplanetary disc contributes all the other elements, such as deuterium. From the abundance of deuterium accreted at the start of planetary formation, we can estimate the corresponding amount of carbon accreted at the same time and quantify the additional carbon accretion required once the protoplanetary disc phase is over. This additional carbon accretion is smaller than the C/H currently observed, and it is this value that must be compared with the results of our simulations.

Consider an arbitrary atmospheric volume $V_{\rm atm}$. This volume contains $N_H^f$ atoms of hydrogen. Of these $N_H^f$ atoms, $N_H^{\rm proto}$ come from the primordial gas, and $N_H^{\rm ppd}$ from the accretion of solid bodies during the protoplanetary disc phase. A fraction of this hydrogen is deuterium, but the ratio $\frac{D}{H}$ is expected to be different for the hydrogen of the primordial nebula $\left. \frac{D}{H} \right|_{\rm proto}$, the solid bodies $\left. \frac{D}{H} \right|_{\rm solids}$, and the final atmospheres $\left. \frac{D}{H} \right|_{\rm atm}$. Knowing their respective values, we can obtain an estimate of $N_H^{\rm ppd}$ by calculating the number of deuterium atoms as follows

\begin{equation}
   \left. \frac{D}{H} \right|_{\rm atm} N_H^f = \left. \frac{D}{H} \right|_{\rm solids} N_H^{\rm ppd} + \left. \frac{D}{H} \right|_{\rm proto} N_H^{\rm proto},
\end{equation}

\noindent where we assume $\frac{N_H^f}{N_H^{\rm proto}} \sim 1$ because the atmospheres of Uranus and Neptune are mainly composed of primordial gas despite their high metallicities (i.e., the extra amount of hydrogen from solid materials is negligible). This assumption will help simplify our equations further to reach the final contribution of early accretion to carbon enrichment.

The solids are composed of water and refractory materials with a mass ratio of $\chi = \frac{M_{H_2O}}{M_{\rm refr}} \sim 1$~\citep{marschall_refractory--ice_2025}, with $M_{\rm refr}$ the mass of refractory materials and $M_{\rm H_2O}$ the water ice mass. $N_H^{\rm ppd}$ can be separated into two distinct contributions that are $N_H^{\rm ppd, w}$ due to water ice (i.e. leading to no carbon enrichment), and $N_H^{\rm ppd, refr}$ due to refractory materials, which produces carbon enrichment.
Let $r_{ref}^w = \frac{N_H^{\rm ppd, w}}{N_H^{\rm ppd, refr}}$, then $N_H^{\rm ppd}=N_H^{\rm ppd, refr} (1+r_{ref}^w)$.
Assuming that the $\frac{C}{H}$ ratio in refractory materials is $\left. \frac{C}{H}\right|_{\rm refr} \sim 1$~\citep{isnard_hc_2019, hanni_identification_2022}, we can finally get an estimate of the atmospheric carbon enrichment during the protoplanetary disc phase using that

\begin{equation}
    N_{C}^{\rm ppd} = \left.\frac{C}{H}\right|_{\rm refr} N_H^{\rm ppd, refr},
\end{equation}
\noindent which can be used to estimate the associated $\frac{C}{H}$ ratio

\begin{equation}\label{finalcsurhfromdsurh}
    \left. \frac{C}{H} \right|_{\rm ppd} \sim \frac{\left. \frac{C}{H} \right|_{\rm refr}}{1 + r_{ref}^{w}} \frac{\left.\frac{D}{H} \right|_{\rm atm} - \left.\frac{D}{H} \right|_{\rm proto}}{\left.\frac{D}{H} \right|_{\rm solids}}.
\end{equation}

The measurements of the D/H ratios of Uranus and Neptune lead to $\left. \frac{D}{H}\right|_{\rm atm} = (4.4 \pm 0.4) \times 10^{-5}$ and $(4.1 \pm 0.4) \times 10^{-5}$, respectively~\citep{guillot_giant_2023}. The protosolar D/H ratio is equal to $\left. \frac{D}{H}\right|_{\rm proto} = (1.67 \pm 0.25) \times 10^{-5}$~\citep{asplund_chemical_2021}. However, the mean D/H in accreted solids is not very well constrained with comet-like objects having D/H ratios between $10^{-4}$ and $10^{-3}$~\citep{altwegg_67pchuryumov-gerasimenko_2015}.
Let us use the direct measurement of the solid $\frac{D}{H}$ ratios for the comet 67P/Churyumov–Gerasimenko as a reasonable assumption. In-situ measurements lead to $\left.\frac{D}{H} \right|_{\rm solids}$ equal to $1.57 \pm 0.54 \times 10^{-3}$~\citep{paquette_dh_2021}.

Now, let us estimate the value of $r_{ref}^w=\frac{N_H^{\rm ppd, w}}{N_H^{\rm ppd, refr}}$ to finally obtain the C/H contribution from solid accretion.
Let $N_H^{H_2O}$ and $N_H^{\rm ref}$ be the number of hydrogen atoms for a molecule of water and a molecule of refractory, $\mu_{H_2O}$, $\mu_{\rm ref}$ the average molecular masses of water and refractory, respectively, and $\rho_{H_2O}$, $\rho_{\rm ref}$ their respective densities. We can then write

\begin{equation}
    r_{ref}^w = \frac{N_H^{H_2O} \frac{\rho_{H_2O}}{\mu_{H_2O}}}{N_H^{\rm ref} \frac{\rho_{\rm ref}}{\mu_{\rm ref}}}=\frac{N_H^{H_2O}}{N_H^{\rm ref}}\frac{\mu_{\rm ref}}{\mu_{H_2O}} \chi^{-1},
\end{equation}
\noindent with $N_H^{H_2O}=2$ and $\mu_{H_2O}=18$ proton mass. We estimate $N_H^{\rm ref}$ and $\mu_{\rm ref}$ from the average solid composition C$_1$H$_{1.56}$O$_{0.134}$N$_{0.046}$S$_{0.017}$ measured by ROSINA for comet the 67P/C-G~\citep[i.e. $\mu_{\rm ref}=16.62$ m$_p$ and $N_H^{\rm ref}=1.56$,][]{hanni_identification_2022}.

Finally, from Eq.~\ref{finalcsurhfromdsurh}, we can compute the carbon enrichment from accreted solids in the protoplanetary disc phase to be $\left[\frac{C}{H}\right]_{\rm ppd} \sim 18$ for both planets, which is smaller than the observed values for Uranus and Neptune closer to 50-80 times the protosolar abundance~\citep{guillot_giant_2023}. From the difference, we can estimate the potential late gas extra carbon enrichment for Uranus to be $\left[\frac{C}{H}\right]_{\rm KB}^{U} \sim 40 \pm 15$ and $\left[\frac{C}{H}\right]_{\rm KB}^{N} \sim 55 \pm 18$ for Neptune, which is slightly higher than the values derived with the S/H method, though of the same order of magnitude. We note that in this case, we find that $\left[\frac{C}{H}\right]_{\rm KB}^{U}>25$ and $\left[\frac{C}{H}\right]_{\rm KB}^{N}>37$ assuming the largest value in our range of $\left.\frac{D}{H} \right|_{\rm solids}$ values comprised between $10^{-4}$ and $10^{-3}$. This may indicate that late gas is indeed necessary if the extent of average values of $\left.\frac{D}{H} \right|_{\rm solids}$ is indeed within that range.

\subsection{The light (present-day) Kuiper Belt (LKB)} \label{subsection_light_belt}

In the case of debris discs, it is anticipated that the viscous parameter will be somewhat elevated in comparison to that observed in protoplanetary discs, with a likely range of values between $\alpha=10^{-3}$ to $10^{-1}$ due to higher ionisation fraction in the former~\citep{kral_magnetorotational_2016, cui_dynamics_2024}. The fiducial simulation (LKB-fid) has an $\alpha$ value of $10^{-3}$, however, a belt with a higher $\alpha$ was also run to check the effect of this parameter (see Table~\ref{table:1}).
In the LKB-fid simulation, we fixed the accretion efficiency to be 0.5; however, we also tested an efficiency of 0.1 and 0.8 (see Tab.~\ref{table:1}).

\begin{table}
\caption{List of the LKB simulations.}
\label{table:1}
\centering
\begin{tabular}{c c c c}
\hline\hline
Simulation & $\alpha$ \tablefootmark{a} & f$_{accr}$ \tablefootmark{b} & M$_{belt} \ [M_\oplus]$ \tablefootmark{c}\\
\hline
   LKB-fid & $10^{-3}$ & 0.5 & 0.1 \\
   LKB-2 & $10^{-2}$ & 0.5 & 0.1 \\
   LKB-3 & $10^{-3}$ & 0.1 & 0.1 \\
   LKB-4 & $10^{-3}$ & 0.8 & 0.1 \\
\hline
\end{tabular}
\newline
\tablefoottext{a}{Diffusion coefficient \citep{shakura_black_1973}.}
\tablefoottext{b}{Accretion efficiency defined in equation \ref{equation_f_accr}.}
\tablefoottext{c}{Initial belt mass in earth masses.}
\end{table}

\begin{figure}[tbh]
\begin{centering}
\includegraphics[scale=0.55]{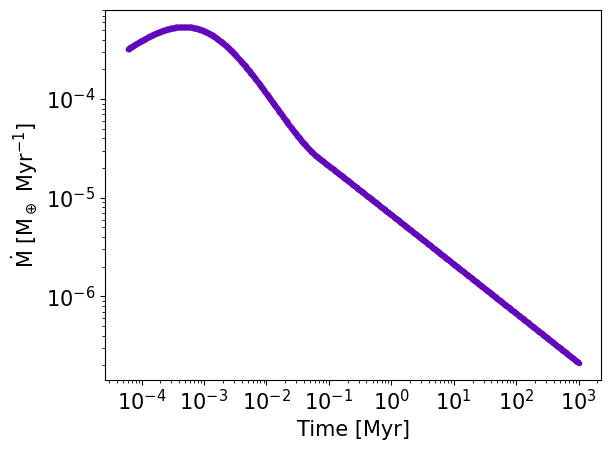}
\end{centering}

\caption{\label{mdotb}CO mass production rate in the belt as a function of time for all LKB (light KB) simulations.}

\end{figure}

\begin{figure}[tbh]
\begin{centering}
\begin{minipage}[c]{0.5\textwidth}
\includegraphics[scale=0.52]{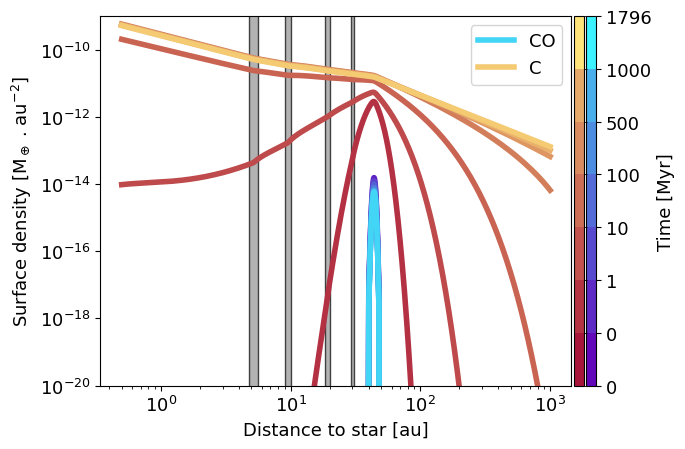}
\end{minipage}\hfill
\end{centering}
\caption{Surface density of CO (in variations of blue) and neutral carbon (in variations of orange) at different times for the LKB-fid simulation. The color goes from darker to brighter for increasing time as indicated by the color bar. The gray vertical bars represent the planets' accretion zones (from left to right: Jupiter, Saturn, Uranus and Neptune).}
\label{sigma_LKBfid}
\end{figure}

The gas production rate in the belt is independent of $\alpha$ and $f_{\rm accr}$ but depends on the initial belt mass, which we take equal to its current value of $\sim$0.1 M$_\oplus$ for the present-day KB~\citep{vitense_improved_2012}. Hence, the quantity of CO gas released in the belt is equal for all LKB's simulations and never exceeds $6 \times 10^{-4} \ \mathrm{M_\oplus / Myr}$ as seen in Figure~\ref{mdotb}. We find that CO is not shielded from photodissociation for the $\alpha$ values that were explored in our set of LKB simulations (e.g., see Figure~\ref{sigma_LKBfid} for the LKB-fid simulation where CO remains co-located with the KB). As a consequence CO gets photodissociated before being able to spread far outside its birthplace (the KB) and the gas accreted by planets is composed of neutral carbon and atomic oxygen.

\begin{figure}[htbp]
\begin{centering}
\includegraphics[scale=0.57]{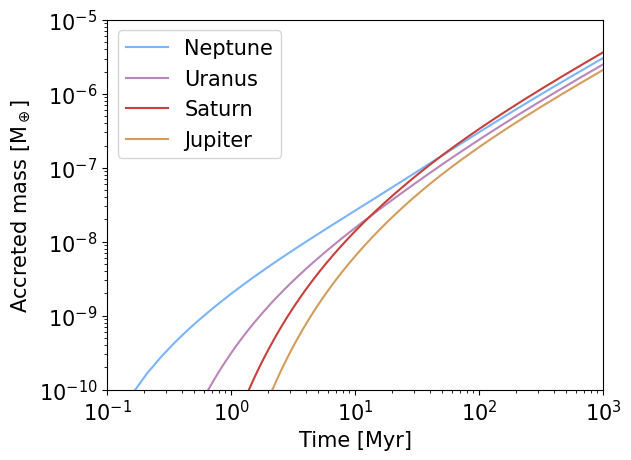}
\end{centering}
\caption{Neutral carbon cumulated accreted mass for each planet as a function of time for the LKB-fid simulation.}
\label{m_accr_LKB_fid}
\end{figure}

In Fig.~\ref{mdotb}, we can see that there is a peak of gas production rate initially at $t\sim500$ yr. This is because all the small grains lose their ices very rapidly. However, most of the ice mass is in the large solids and the majority of it is released over large timescales as can be seen in both Figs.~\ref{sigma_LKBfid} and~\ref{m_accr_LKB_fid}.
The rate of gas production throughout most of the simulations is around $10^{-6}$ M$_{\oplus} / \mathrm{Myr}$ between 10-100 Myr and decreases slowly over time. This is because heat takes time to diffuse in the largest bodies (the thermal conductivity $K$ value is much lower than in asteroids for example), where most of the mass is located given the size distribution assumed. Hence, there is still a significant amount of ice in the belt after 1 Gyr (see Figure~\ref{sigma_LKBfid}).

Once produced, the gas viscously evolves and when the steady state is reached after roughly 10 Myr (i.e. when gas had time to spread both inwards and outwards), we find that there are no significant differences in the gas mass accreted by the different giant planets as a function of time (see Figure~\ref{m_accr_LKB_fid}). This might appear counter-intuitive because Jupiter and Saturn being located farther out from the KB than the icy giants, one may expect that they receive less gas as they may get starved by Uranus and Neptune. However, given their higher masses and corresponding Hill radii, they also have greater incoming mass fluxes (see Eq.~\ref{phiin}).
Hence, the greater degree of carbon enrichment due to late gas accretion for the atmospheres of Uranus and Neptune observed in our simulations (Fig.~\ref{CH_LKB_fid}) over that of Jupiter and Saturn (Fig.~\ref{CH_LKB_fid_giants}) is due to their smaller atmospheric masses (in which carbon can mix) rather than more mass accreted.

Changing $\alpha$ only affects the time to reach steady state. However, the final accreted masses remain similar for all values of $\alpha$ tested (see Figure~\ref{m_accr_LKB}). It is expected because $\alpha$ does not influence the amount of gas mass released from the belt nor its accretion efficiency onto planets.

We do not observe a significant neutral carbon enrichment/depletion due to interspecies diffusion (see Fig.~\ref{sigma_LKBfid}). The C/O ratio of the accreted gas is then always equal to 1 in our simulations, since we assume that only CO ices sublimate. We note that it is unclear whether some CO$_2$ ices may also sublimate in the KB in addition to CO, which would then decrease the gas disc C/O ratio~\citep{de_pra_widespread_2024}.

In Figure~\ref{m_accr_LKB}, we can observe that there is no direct proportionality between the accreted mass and the accretion efficiency $f_{\rm accr}$. The accreted mass for the icy planets in the LKB-fid simulation (with $f_{\rm accr}=0.5$) is about 20 times higher (i.e. much greater than five times) than the accreted mass in LKB-3 (with $f_{\rm accr}=0.1$). As can be seen in the same figure, the icy planets demonstrate a greater capacity to accrete mass compared to giants when the efficiency of the accretion process is high. This is because, in the case of a higher accretion efficiency, the gas giants are deprived of gas that is accreted onto the icy planets, which results in a lower accreted mass than in the fiducial simulation.

Finally, regardless of the circumstances, we find that for the LKB simulations the accreted masses on both Uranus and Neptune turn into a small C/H value when compared to the observed values. For instance, for our LKB-fid simulation, we can anticipate a maximum of one percent of the protosolar C/H ratio in the atmospheres of icy planets due to accretion of late gas from a light Kuiper Belt (see Figures~\ref{CH_LKB_fid}). The results obtained for the other simulations are analogous, given the comparable accretion rates (See table~\ref{tab:LKB_results}).

\subsection{The heavy (primordial) Kuiper belt (HKB)}\label{HKBpart}

The primordial Kuiper belt is expected to have been much more massive than the current KB with a much higher gas production rate than that predicted in the previous subsection. To evaluate the efficiency of gas accretion onto the giants from gas released in the young primordial KB, a series of simulations were conducted. The predictions for the accretion of gas onto the icy giants will vary depending on the assumed primordial KB model (as described in subsection~\ref{subsection_belt_model}), with the most important parameters being the initial mass of the KB (5, 20, 50 Earth masses), its lifetime (10 to 100 Myr), and the system's architecture (compact or extended). The different setups of our heavy KB (HKB) simulations are listed in Table~\ref{table:simulations_HKB}. The final $\left[\frac{C}{H}\right]$ for all simulations are summarized in Table~\ref{tab:HKB_results}.

\begin{table}
\caption{List of the HKB simulations.}
\label{table:simulations_HKB}
\centering
\begin{tabular}{c c c c c c}
\hline\hline
HKB \tablefootmark{a} & f$_{accr}$ \tablefootmark{b} & M$_{belt}^\mathrm{0}$ $[M_\oplus]$ \tablefootmark{c} & system \tablefootmark{d} & $t_{\rm ev}$ \tablefootmark{e}& $t_{\rm fold}$ \tablefootmark{f}\\
\hline
   fid & 0.5 & 50 & c & 10 & 10\\
   2 & 0.1 & 50 & c & 10 & 10\\
   3 & 0.5 & 20 & c & 10 & 10\\
   4 & 0.5 & 5 & e & 10 & 10 \\
   5 & 0.5 & 5 & e & 100 & 100 \\
\hline
\end{tabular}
\newline
\tablefoottext{a}{HKB simulations names}
\tablefoottext{b}{Accretion efficiency defined in equation \ref{equation_f_accr}.}
\tablefoottext{c}{Initial belt mass in earth masses.}
\tablefoottext{d}{Positions of the planets. In the compact configuration (c in the table) Neptune is initially much closer to the Sun than in the extended configuration (e in the table). The belt is at the same place for all HKB simulations, i.e. closer to the Sun than the present-day Kuiper belt.}
\tablefoottext{e}{Starting time of the depletion by Neptune's outer migration in Myr.}
\tablefoottext{f}{Exponential dissipation of the belt mass over an e-folding time $t_{\rm fold}$ (in Myr in the table) starting at $t_{\rm ev}$ .}
\end{table}

\begin{table}[]
    \caption{List of the Minimal and Maximal estimations for all HKB simulations of the atmospheric [C/H] ratios for Jupiter (J), Saturn, (S), Uranus (U), and Neptune (N).}
    \label{tab:HKB_results}
    \centering
    \begin{tabular}{c|c|c|c|c}
    \hline\hline
       HKB &  J & S & U & N \\
    \hline
    fid & $0.14$ -- $0.16$
    & $0.93$ -- $2.0$
    & $12.9$ -- $33.3$
    & $20$ -- $58.3$ \\
    2 & $0.09$ -- $0.09$
    & $0.34$ -- $0.73$
    & $3.6$ -- $9.4$
    & $5.2$ -- $13.1$ \\
    3 & $0.04$ -- $0.049$
    & $0.27$ -- $0.57$
    & $3.7$ -- $9.5$
    & $6.6$ -- $16.7$ \\
    4 & $0.007$ -- $0.008$
    & $0.04$ -- $0.10$
    & $0.6$ -- $1.6$
    & $1.0$ -- $2.6$ \\
    5 & $0.012$ -- $0.013$
    & $0.07$ -- $0.17$
    & $1.1$ -- $2.8$
    & $1.8$ -- $4.9$ \\

    \hline
    \end{tabular}
\end{table}

\subsubsection{The compact architecture}

The compact architecture represents a scenario akin to that of the Nice model~\citep{tsiganis_origin_2005, levison_late_2011, griveaud_solar_2024}. The planets begin in a compact configuration, with a substantially massive belt of 20 to 50 M$_\oplus$. The outer planets migrate outward, resulting in the depletion of the primordial belt.
This is the most comprehensive model to date of the evolution of the young solar system, capable of reproducing both the current configuration of planets and the present-day Kuiper belt. Accordingly, the characteristics of our HKB-fid, 2, and 3, simulations have been selected to align with this architecture (see subsection~\ref{subsection_belt_model}).

\begin{figure}[h]
\begin{centering}
\begin{minipage}[c]{0.49\textwidth}
\includegraphics[scale=0.53]{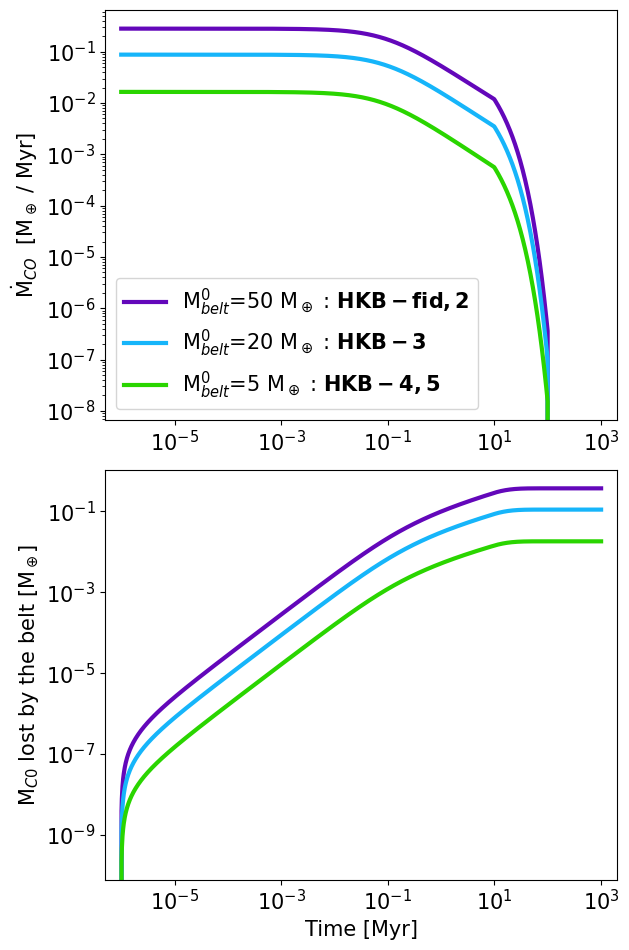}
\end{minipage}\hfill
\end{centering}
\caption{{\it Top}: CO mass production rate in the belt for the HKB simulations.
{\it Bottom}: Total mass of CO gas released in the belt for the HKB simulations. The initial mass of 5 M$_\oplus$ (HKB-4, 5) is for the extended configuration, whereas the initial masses of 20 (HKB-3) and 50 M$_\oplus$ (HKB-fid, 2) are for the compact configuration.}
\label{m_dot_NM}
\end{figure}

In our fiducial simulation, probably the most realistic, we have selected a relatively substantial primordial belt of 50 Earth masses in agreement with recent work showing that assuming a more realistic low viscosity for the protoplanetary disc leads to a more massive primordial KB~\citep{griveaud_solar_2024}.
In Figure~\ref{m_dot_NM} (bottom), we can see that the HKB-fid simulation lead to a total CO mass released greater than 0.1 M$_\oplus$ in roughly 10 Myr. On the same figure (top), we see that the gas production rate reaches values greater than 0.1 M$_\oplus$/Myr, i.e. orders of magnitude higher than for the LKB simulations. For the HKB-3 simulation with a belt mass of 20 earth masses, the total CO mass released and gas production rates are lower but of the same order of magnitude.

\begin{figure}[h]
\begin{centering}
\begin{minipage}[c]{0.49\textwidth}
\includegraphics[scale=0.52]{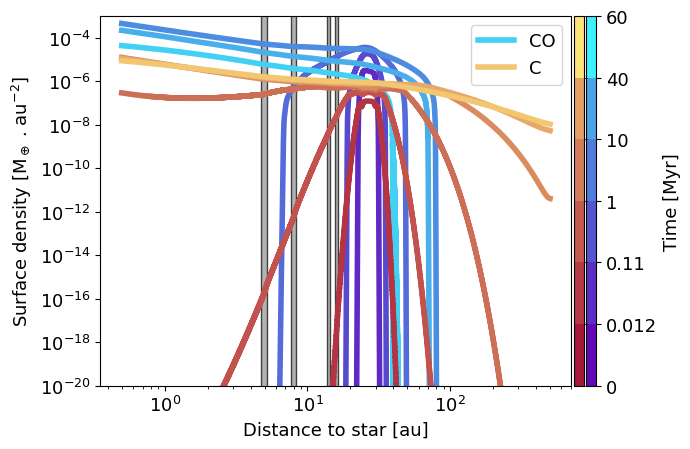}
\end{minipage}\hfill
\end{centering}
\caption{Surface density of CO (in variations of blue) and neutral carbon (in variations of orange) at different times for the HKB-fid simulation. The color goes from darker to brighter for increasing time as indicated by the color bar. The gray vertical bars indicate the planets' accretion zones (from left to right: Jupiter, Saturn, Uranus and Neptune).}
\label{sigma_NM}
\end{figure}

In Figure~\ref{sigma_NM}, we see that for the HKB-fid, 2, 3 simulations, CO gets shielded from the impinging interstellar radiation. Hence, CO has time to spread viscously within the system and it can reach the giants in molecular form. Therefore, the accreted gas onto the icy planets is mostly made up of CO rather than atomic carbon and oxygen as was the case for the LKB simulations explored in the previous section. However, CO is not stable in the upper layers of the atmospheres of Uranus and Neptune, and it probably reacts with the primordial hydrogen in the atmospheres to create CH$_4$ and H$_2$O~\citep{atreya_deep_2020}. The extra carbon brought via late gas could then be seen when observing the C/H ratio in the atmospheres. As discussed later, we note that the O/H ratio is not very informative because water is not only confined to the atmospheres and the interpretation of the O/H ratio is more complex.

For the HKB simulations, the $\alpha$ value was fixed to $10^{-3}$. We do not vary it as it does not have an important impact on the final predictions of the captured gas mass by planets as previously explored for the LKB. However, there is still plenty of gas at the onset of Neptune's depletion of the primordial belt (second phase of the simulation, see Figure~\ref{sigma_NM}). We do not model the exact planet's dynamic crossing the disc during its outward migration, and the accretion rates obtained after this time might then be over or underestimated. However, to account for that, we explore the final masses we obtain for different belt lifetimes going from 10 to 100 Myr, which can lead to a factor 2 difference in total mass captured onto planets.

\begin{figure*}[h]
\begin{centering}
\includegraphics[scale=0.38]{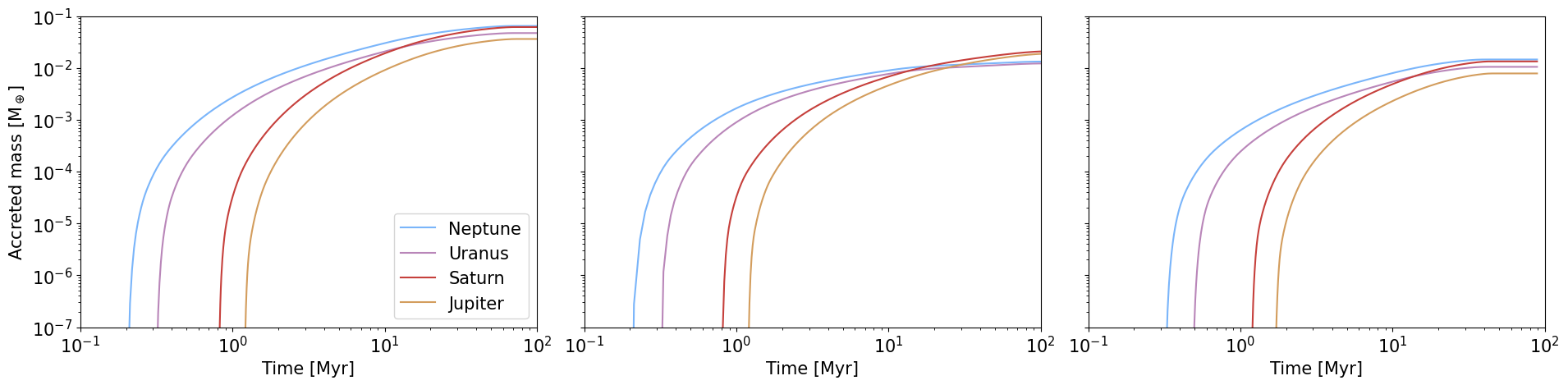}
\caption{CO accreted mass for each planet as a function of time for the compact configuration of HKB simulations (HKB-fid left, HKB-2 middle, HKB-3 right).}
\end{centering}
\label{accreted_mass_HKB}
\end{figure*}

Figure~\ref{accreted_mass_HKB} shows the CO mass captured onto the 4 giants for our HKB-fid, 2, 3 simulations. Note that the HKB-fid and HKB-2 only differ by their $f_{\rm accr}$ values of 0.5 and 0.1, respectively, and the HKB-3 tests a less massive belt of 20 M$_\oplus$. Figure~\ref{accreted_mass_HKB} shows that the CO mass accreted onto icy giants is greater than $10^{-2}$ M$_\oplus$, which may be significant when injected into the atmospheres of Neptune and Uranus. Indeed, assuming that the accreted CO only mixes in the atmospheres of Uranus~\citep[1.25-3.5 M$_\oplus$,][]{guillot_giant_2023}, and Neptune (1.6-4.15 M$_\oplus$), we can easily compute the corresponding [C/H] ratio.

\begin{figure}[tbh]
\begin{centering}
\begin{minipage}[c]{0.45\textwidth}
\includegraphics[scale=0.6]{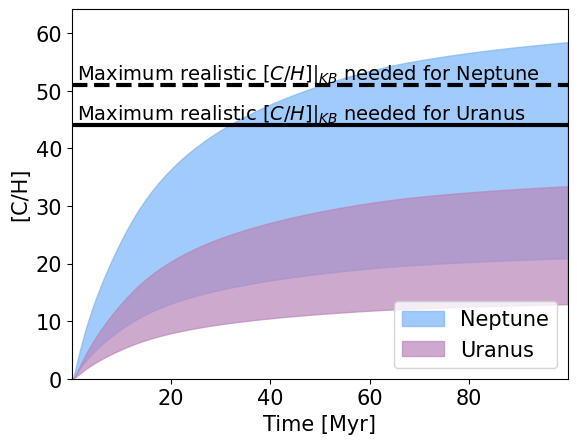}
\end{minipage}\hfill
\end{centering}
\caption{[C/H] for the HKB-fid simulation as a function of time. The filled areas correspond to predictions for Uranus (purple) and Neptune (blue). The uncertainties are shown via the extension of the filled areas and are due to uncertainties on the respective atmospheric masses. The solid and dashed black lines are the maximum estimation of the gas disc contribution obtained from observations for Uranus and Neptune (see subsection \ref{soverh}).}
\label{CH_NM}
\end{figure}

Figure~\ref{CH_NM} illustrates that after 10 Myr of late gas accretion, the [C/H] ratio of Uranus reaches 5 to 13, while for Neptune it reaches 9.5 to 25. These are the values only accounting for late gas accretion and assuming that no enhanced primordial metallicity was present. We note that if the gas disc survives 100 Myr instead, the resulting [C/H] ratio is 13.5 to 33 for Uranus and 20 to 58 for Neptune. We remind the reader that observations show that [C/H] is of $44-74$ and $55-92$ for Uranus and Neptune~\citep{atreya_deep_2020}, i.e. our fiducial model may not be far from potentially explain it all by itself.

However, some primordial enhancement of C/H may be present due to the planet formation process (e.g. pebble accretion) and a smaller C/H may be needed to account for observations as explained in the beginning of the result section. To account for that, in Figure~\ref{CH_NM} we plot two horizontal lines showing the maximum extra carbon enrichment needed to explain current observations with the primordial enrichment calculated using measurements of S/H. Our calculations show that Uranus should accrete on average an extra [C/H]  of 24 and Neptune 21. However, given uncertainties, we can only reliably derive upper values of extra [C/H] of 51 for Neptune and 44 for Uranus (see section~\ref{section_discussion} for more details), which are plotted on Figure~\ref{CH_NM}. We conclude that the fiducial simulation is able to explain both observed [C/H] values of Uranus and Neptune, with most of the metallicity enrichment coming from accretion of late gas.

The HKB-fid simulation represents the most favourable case for the accretion of carbon-rich gas onto icy planets, characterized by the most massive belt and the highest accretion efficiency.
The HKB-2 simulation was run with an accretion efficiency five times smaller than the HKB fiducial value to test the impact of $f_{\rm accr}$ on our results. It leads to a C/H ratio lower than that of the fiducial simulation, by a factor of between 1.4 and 9.2 for Uranus and 1.5 and 11.2 for Neptune, after 100 million years.

\begin{figure}[tbh]
\begin{centering}
\begin{minipage}[c]{0.45\textwidth}
\includegraphics[scale=0.55]{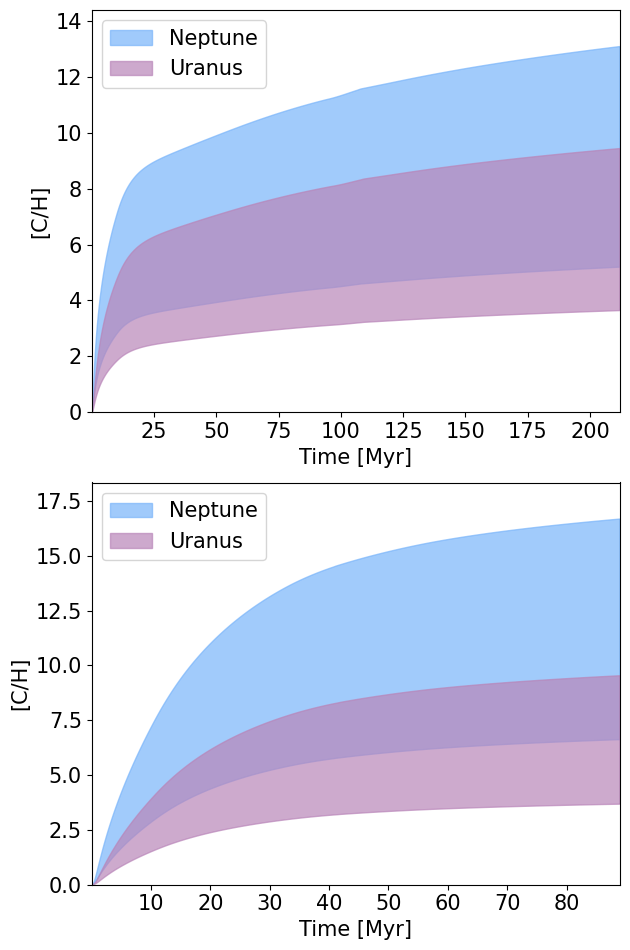}
\end{minipage}\hfill
\end{centering}
\caption{[C/H] for the simulation HKB-2 (top) and HKB-3 (bottom) as a function of time.}
\label{CH_NM2}
\end{figure}

The HKB-3 simulation was run with a less massive belt because even though recent Nice model simulations~\citep{griveaud_solar_2024} suggest it may have reached 50 M$_\oplus$, the collisional models of the belt by~\cite{bottke_collisional_2023} indicate the possibility of a slightly lighter belt of about 20 M$_\oplus$. We find results very similar to the HKB-2 simulation as seen in Figure~\ref{CH_NM2}. The C/H values reached by both simulations are still within the right range to explain plausible extra carbon-enrichment from late gas in Uranus and Neptune, but although significant, it would become a minor source compared to primordial enrichment.

Furthermore, our model also allows us to assess the quantity of carbon-rich gas that can accumulate on the gas giants Jupiter and Saturn. Despite the similarity in absolute accreted mass, the atmospheric mass in which the accreted gas will mix is significantly higher. We account for the effect of dilution of the solid core for Jupiter and Saturn that can change the atmospheric mass into which carbon can mix with accretion. We consider lower limits for the atmospheric mass using a diffuse core of 60 M$_\oplus$ for both Jupiter and Saturn, and upper limits of 20-30 M$_\oplus$ for the solid cores of Saturn and Jupiter, respectively.

\begin{figure}
\begin{centering}
\includegraphics[scale=0.6]{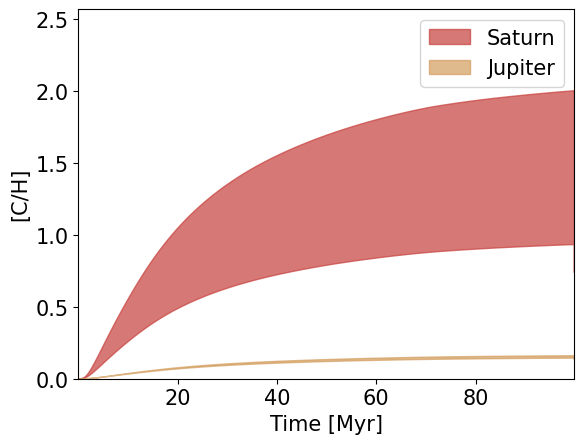}
\end{centering}
\caption{[C/H] for the simulation HKB-fid as a function of time for Jupiter and Saturn. The uncertainties are shown by the extent of the filled-in areas and are due to the uncertainties over the respective atmospheric mass in which the CO can mix (i.e. between a diluted 60 M$_\oplus$ diffuse core for both planets and a 20, 30 M$_\oplus$ solid core for Saturn and Jupiter, respectively). }
\label{CH_NM_giants}
\end{figure}

The contribution of late gas released in the primordial Kuiper belt to the [C/H] ratio of Jupiter and Saturn is, at most, of 0.16 and 2.0, respectively (see Figure~\ref{CH_NM_giants}). If we consider a dilute core of 50 $\%$ of the total mass for Jupiter the [C/H] is at most 0.26. However, the observed [C/H] ratios are also much lower for Jupiter and Saturn~\citep[respectively 3.56 $\pm$ 0.86 and 7.50 $\pm$ 0.32]{guillot_giant_2023}, hence this contribution might also be non-negligible at least for Saturn. Indeed, the upper limits for the contribution of late gas (calculated using the S/H ratios as a tracer of accretion from planet formation) are 2.19 and 4.7 for Jupiter and Saturn, respectively.

We conclude that our compact configuration simulations (mimicking the Nice model) provide very compelling results and show that late gas could be the major contributor to the observed C/H values in both Uranus and Neptune in the best-case scenario, or at least lead to non-negligible enhancement in the worst-case scenario, which is still needed to be accounted for. We will discuss the consequences in a broader context in section~\ref{section_discussion}.

\subsubsection{The extended architecture}

In the compact architecture context, the planets and KB need to start in a compact configuration with a massive planetary belt to end up with the present-day solar system configuration at the end of the system's evolution. However, some mechanisms allow the planets to migrate earlier, directly in the protoplanetary disc phase as, for instance, the ``Rebound'' model by \citet{liu_early_2022}. In this case, once the protoplanetary disc fully dissipates, the system has its planets and belt at locations very similar to the present-day solar system, i.e. more extended than in the compact configuration. In this context, the belt mass can be smaller (e.g. 5 M$_\oplus$) while still allowing to reach the current Solar System architecture~\citep{liu_early_2022}.

We ran two simulations mimicking the extended architecture approach with a 5 M$_\oplus$ primordial belt and varying belt lifetimes (10 Myr for HKB-4 and 100 Myr for HKB-5, see Table~\ref{table:simulations_HKB}). The latter is important because it is not yet clear what is the exact dynamics of belts in the extended configuration and it may take longer to deplete that in the compact configuration~\citep{liu_early_2022}.

\begin{figure}[tbh]
\begin{centering}
\includegraphics[scale=0.55]{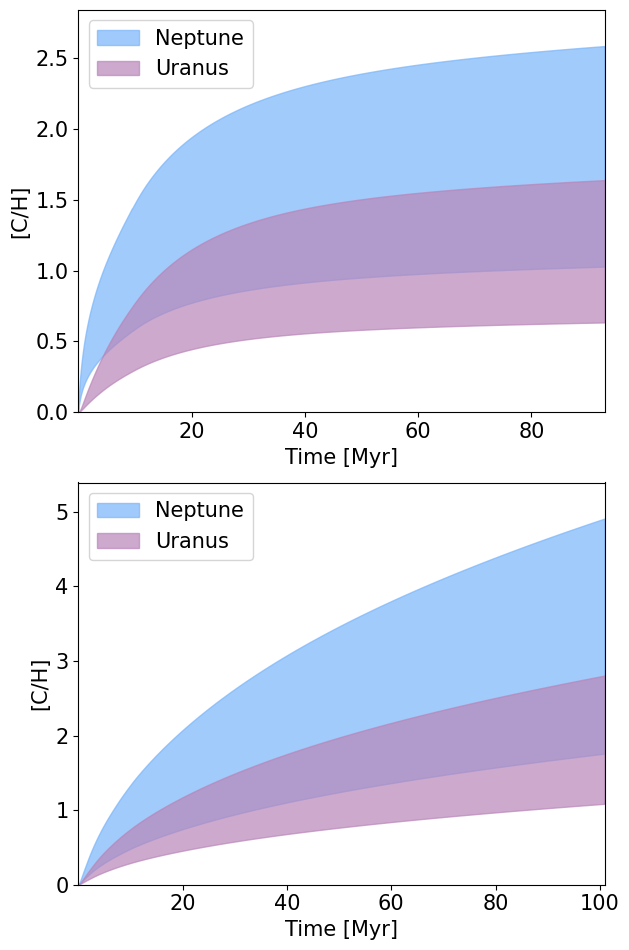}
\end{centering}
\caption{[C/H] for the simulation HKB-4 (top) and HKB-5 (bottom) as a function of time.}
\label{CH_HKB_extended}
\end{figure}

Figure~\ref{CH_HKB_extended} shows the C/H predictions for late gas capture on Uranus and Neptune for the HKB-4 and HKB-5 simulations. As expected, because of the lower initial belt mass, we end up with smaller C/H ratios than the HKB-fid simulation. In this case, irrespective of the lifetime of the belt, the contribution of late gas to the atmospheric carbon enrichment for Uranus and Neptune is small compared to observed ratios and of order a few, though not negligible.

We conclude that the extended configuration (mimicking other models than the Nice model) does not lead to a significant contribution to the metallicity of Uranus and Neptune. We note, however, that if it is confirmed by future observations that extra metallicity is needed to explain the C/H in both ice giants, our work could be used as an argument in disfavor of those extended models, thus reinforcing the Nice model.

\section{Discussion} \label{section_discussion}

\subsection{The O/H ratio}
Together with carbon, some oxygen will be accreted onto the planets, either in the form of CO, or directly as atomic oxygen. We then naively expect the atmospheres of Uranus and Neptune to be enriched in oxygen compared to solar abundances. However, water condenses deep into the atmosphere~\citep[e.g.,][]{cano_amoros_h2h2o_2024}, and we only have an indirect constraint on the O/H abundance through the detection of CO, and then using models of chemical reaction rates converting CO into CH$_4$ and accounting for upward transport of this CO~\citep{lellouch_dual_2005,cavalie_first_2014, venot_new_2020}. Therefore no reliable O/H measurement can be used to estimate the amount of solids accreted during planet formation or the contribution from late gas, and we do not explore this aspect further though it may be different for exoplanets with different temperatures as we discuss later in subsection~\ref{extra}.

\subsection{The primordial belt model to explain the metallicity of Uranus and Neptune}

\subsubsection{The gas viscosity}

The $\alpha$ viscosity in debris discs is expected to be rather high compared to protoplanetary discs with values that can reach $10^{-3}-0.1$ both because of the magnetorotational instability~\citep{kral_magnetorotational_2016} or other instabilities, or even via the molecular viscosity which may not be negligible in debris disc conditions~\citep{cui_dynamics_2024}.
    It is thus expected that the gas will spread from the KB to Neptune in $t_{\rm vis}=\Delta R^2/\nu \sim $ 0.3 to 30 Myr for $\alpha$=0.1 to $10^{-3}$.
    As one can see the viscous timescale decreases linearly with $\alpha$ and the only difference in our simulations will thus be the time to reach planets but the accreted mass will remain the same as shown by the LKB-2 simulation and explained in further detail in \citet{kral_impact-free_2024}. Hence, in the long term, we do not expect $\alpha$ to play a major role for the final accreted mass of gas onto planets.

We note that one difference it could make is on the composition of the gas disc. Indeed, if $\alpha$ is small, the gas disc will spread more slowly and become denser~\citep{kral_magnetorotational_2016, kral_imaging_2019} and some shielding may appear (e.g. from carbon or CO). However, it is not very relevant for our study because the final quantity of C and O accreted into the atmospheres of Neptune and Uranus will not change whether it is captured as CO or C+O. This gas mass will then be turned into CH$_4$ and water into the atmospheres of Uranus and Neptune as explained in subsection~\ref{HKBpart}.

\subsubsection{Effects of the belt's mass}

The gas production rate we use is based on the solid mass loss rate in the KB, which we estimate using Eq.~38 in \citet{lohne_long-term_2008}. This collisional model accounts for varying slopes in the size distribution with e.g, a primordial slope (for bodies that had not had time to collide significantly yet), and then two more slopes to account for the gravity and strength regimes, respectively. We note that even though simplified one-slope collisional models show that the mass loss rate scales as the square of the belt mass, it is indeed more complicated in our case. Numerical calculations show for instance that the HKB-3 belt of 20 M$_\oplus$ leads to less than 2 / 5 of the gas mass produced by the HKB-fid belt of 50 M$_\oplus$, due to this non-linearity in our model.
As the belt mass increases, the collision timescale decreases as it typically scales with the inverse of total belt mass~\citep{wyatt_evolution_2008}. Increasing the belt mass leads to more collisions, faster, and thus more gas produced when the system is younger. This explains that models with the less massive belts initially (e.g. HKB-5) take longer to release their gas content (e.g. see Fig.~\ref{CH_HKB_extended}).

It is therefore essential to pay attention to the exact size distributions and belt mass given how they may change the final gas production rates.

The outer planets might dynamically heat the belt, especially given the potential outward migration of Neptune~\citep{kenyon_collisional_2004,nesvorny_dynamical_2018}. Therefore, to account for this effect, we assume a mean eccentricity of 0.25 for the planetesimals in the belt. The rate of mass production in such a belt in collisional evolution is typically proportional to $e^{5/3}$~\citep{wyatt_evolution_2008}. Thus, a colder belt with a typical eccentricity of 0.1 will produce approximately 5 times less gas than assumed here. A belt with very high dynamical heating and a typical eccentricity of 0.5 will produce approximately 3 times more gas than in our fiducial model.
We note that this collisional model can explain most observations~\citep{kral_predictions_2017, kral_imaging_2019} but it would be even more satisfying to develop a physical model able to predict directly the amount of gas released from physical mechanisms. However, more data are needed and this is work left for the next decade, which will allow us to refine our predictions.

\subsubsection{Effects of the accretion efficiency}

\begin{figure}[tbh]
\begin{centering}
\includegraphics[scale=0.58]{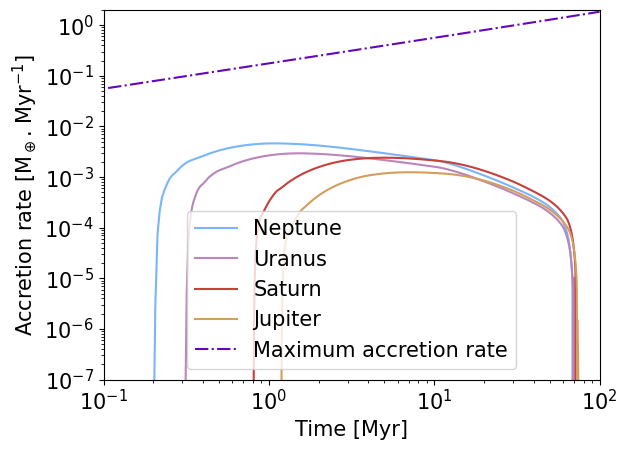}
\end{centering}
\caption{Accretions rates for the HKB-fid simulations compared to the maximum capture rate onto planets due to atmospheric cooling.}
\label{HKB_accretions_rates}
\end{figure}

The accretion efficiency of a planet capturing gas is susceptible to two effects: 1) thermal cooling of the atmosphere~\citep{piso_minimum_2014, piso_minimum_2015, lee_cool_2015, kral_formation_2020}, and 2) hydrodynamical effects~\citep{lubow_gas_2006, mordasini_characterization_2012,tanigawa_distribution_2012,
ormel_hydrodynamics_2015, lambrechts_quasi-static_2019}.
The analytical model developed by \citet{kral_formation_2020} shows that in a debris disc environment cooling is very efficient. If we apply their Eq.~5\footnote{Note that we corrected it by a factor of 4$\pi$.} for Uranus and Neptune, we obtain a maximum theoretical capture rate onto the planets that is higher than accretion rates obtained from the simulations ran in this paper. This means that theoretically, all incoming gas can be captured if only cooling is accounted for as shown in Figure~\ref{HKB_accretions_rates} for the HKB-fid simulation, which leads to the highest accretion rates. We note that the theoretical prediction is analytical and does not account for subtle effects such as the decoupling between the disc and the planet's atmosphere that may happen at low gas densities such as those found in debris disc environments~\citep{mordasini_characterization_2012}. Implementing numerical simulations of cooling in debris disc conditions is not trivial and is left for future work.

As stated earlier, the other physical mechanism that may impact the theoretical accretion rate is the complex hydrodynamical flow.
Previous numerical hydrodynamical simulations in protoplanetary disc conditions show that the accretion efficiency ($f_{\rm accr}$) accounting for hydrodynamic effects is at least 0.1 and most likely higher than 0.5. For instance, 2D hydro simulations by \citet{lubow_gas_2006} provide $f_{\rm accr}$ values between 0.38 and 0.88. We note that we derive their accretion efficiency from the ratio $\dot{M}_i/\dot{M}_p$ they compute. Those results are confirmed by 3D hydrodynamical simulations~\citep[e.g.,][]{schulik_global_2019, li_3d_2023, bailey_systematic_2024}

Gas recycling may be important in protoplanetary discs~\citep{ormel_hydrodynamics_2015, moldenhauer_recycling_2022} but it is tightly linked to the cooling timescale, which is very small in our case and this may not operate. Dedicated hydrodynamic simulations in debris disc conditions would provide further insights concerning the right range of values to use for $f_{\rm accr}$ even though it is most likely of the same order of magnitude as in protoplanetary discs.

We note that gap opening may also impact accretion rates. It is expected that Uranus and Neptune are not massive enough to open a gap~\citep{bergez-casalou_influence_2020} but Saturn and Jupiter would. In this case, accretion rates would be slightly lowered as shown in \citet{bergez-casalou_influence_2020}, which should be accounted for in future hydrodynamical simulations. Another effect, which should be less important would be to consider the migration of Neptune through the gas disc as it accretes. Given that the migration timescale is small compared to the accretion timescale, we do not expect a large change in our conclusions.

For all those reasons, we consider the simulation HKB-2 with an accretion efficiency of 0.1 to most likely provide a lower bound on the atmospheric carbon enrichment for the icy planets. In that worst-case scenario, we find [C/H] values of 9 and 13 (at most) at the end of the simulation for Uranus and Neptune, respectively. We note that after 10 Myr the carbon enrichment is already non-negligible because accretion is much faster at the beginning of the simulation. We consider that it reinforces our conclusion that late gas may have a significant impact on the metallicity of giant ices and may turn out to be important in extrasolar systems as well as discussed in subsection~\ref{extra}.

\subsection{Using our results to constrain the formation model of the Solar System}\label{constrain}

Given the uncertainties on the contribution of early enrichment during planet formation on the observed C/H ratios of Uranus and Neptune, it is not yet possible to give firm conclusions concerning the need for late gas enrichment and hence infer the initial belt mass needed. However, assuming the central values of the derived extra C/H as more likely, we see that an extra [C/H] of $\sim 20$ would be needed for both Uranus and Neptune. Using our model, we see that we need belts with masses greater than 20-30 M$_\oplus$ to be able to reproduce this order of magnitude. If confirmed, it would be an indication that the heavy belt needed in the Nice model to reproduce the current Solar System is needed.

Another constraint that may be relaxed is where the ice giants formed relative to the CO iceline. Indeed, current models postulate that the two planets formed at the CO ice line within the protosolar nebula to be able to explain the high C/H ratio~\citep{irwin_latitudinal_2019, sromovsky_methane_2019}. However, if the carbon contribution comes from late gas, this constraint can be lifted, which could be interesting to consider in future models of planet formation.

\subsection{On the results concerning Saturn and Jupiter}
One of our results is that the carbon enrichment for Saturn is higher than that for Jupiter (see figure~\ref{CH_NM_giants}). This difference between the two planets is greater for higher accretion efficiencies. This is expected because for a higher accretion efficiency, there is less gas arriving at Jupiter, thus reducing its accretion rate. There is also a dilution effect due to the highest atmospheric mass for Jupiter.
In any case, we find that the predicted atmospheric [C/H] ratios due to late gas accretion in the HKB-fid simulation are 2.0 and 0.16 for Saturn and Jupiter, respectively, in agreement with the potential maximum extra C/H needed from late gas accretion of 4.7 and 2.2 (using the S/H ratios as presented in subsection~\ref{soverh}), respectively.
Thus, we find that observations of giant planets are compatible with late gas capture, but given uncertainties on the $\left[\frac{C}{H}\right]_{ppd}$ coming from accreted solids materials during the protoplanetary phase, it is hard to use this information to constrain the disc properties such as its mass or lifetime, similar to Uranus and Neptune as explained in subsection~\ref{constrain}.

\subsection{Late gas accretion onto the inner terrestrial planets}
Up until now, we have focused on the accretion of the four giant planets in the Solar System. However, one notices that in our simulations the gas disc reaches the inner terrestrial planets (see for example Figure~\ref{sigma_NM} for the HKB-fid simulation). Even though we do not include those inner planets directly, we can post-process the amount of CO arriving at the planets' Hill spheres and how much gets accreted using the same technique as described for giant planets. Only accounting for the Earth (i.e. neglecting Mars), we derive a maximum carbon mass that could be accreted onto it in the HKB-fid simulation, that is $\sim 10^{-3}$ M$_\oplus$ in 10 Myr and $\sim 10^{-2}$ M$_\oplus$ in 100 Myr.

However, we do not expect such a high carbon mass accreted for the Earth. In fact, our simulations do not model any gap formation but Jupiter and Saturn are much larger than Uranus and Neptune and they could both open gaps~\citep{bergez-casalou_influence_2020}. Accounting for this effect would reduce the mass flux crossing both Jupiter and Saturn's orbits, thus reducing the potential flux reaching Earth's orbit. We note that the accretion efficiency could also be higher than the assumed 0.5, starving the inner planets, similar to what happens to Jupiter and Saturn in the simulation LKB-4. We also note that in most simulations, gas is accreted onto planets very early, roughly within 10 Myr of protoplanetary disc dissipation where Earth has not yet accreted its final mass~\citep{kleine_tungsten_2017, clement_mars_2018} and is still very hot, thus reducing the accretion efficiency~\citep{piso_minimum_2014, piso_minimum_2015, lee_cool_2015}. Another effect that may be of importance for the Earth is that of giant impacts~\citep{schlichting_atmospheric_2015} including the moon-forming impact~\citep{kleine_tungsten_2017} that may remove part of the accreted CO after it has been accreted. Due to those uncertainties, we decided not to include terrestrial planets in our model and do not attempt to conclude further on late CO gas accretion to terrestrial planets. However, we note that in extrasolar systems with no massive giants, late gas may easily reach those inner planets and alter their atmospheres, which could be an important factor to account for to obtain a proper prediction of their final atmospheric makeup.

\subsection{Application to extrasolar systems}\label{extra}
This paper focuses on the Solar System, however, similar conditions may happen in extrasolar systems. Indeed, it is now accepted that exo-Kuiper belts releasing CO gas may be the norm rather than the exception~\citep{moor_molecular_2017}. Released gas will also spread inwards and outwards of the belt and deliver CO or C+O gas to exoplanets in the system. Our mechanism seems universal and would naturally increase exoplanet metallicities in carbon and oxygen.

\begin{figure}[tbh]
\begin{centering}
\begin{minipage}[c]{0.5\textwidth}
\includegraphics[scale=0.5]{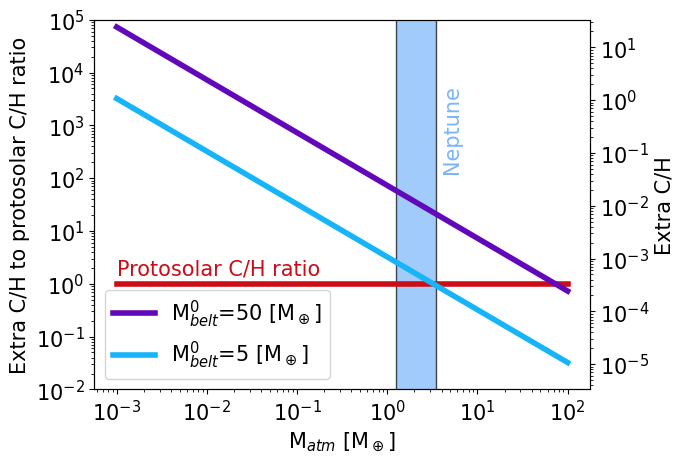}
\end{minipage}\hfill
\end{centering}
\caption{Estimation of the extra C/H due to late gas accretion onto an exoplanet as a function of its atmospheric mass for a belt of respectively 50 M$_\oplus$ and 5 M$_\oplus$ (purple and blue) based on the HKB-fid and HKB-4 simulations. The left axis shows the atmospheric C/H ratio relative to the protosolar abundance, and the right axis shows the absolute atmospheric C/H ratio. The red line represents the protosolar C/H ratio. The blue shaded area represents Neptune's atmospheric mass.}
\label{fig:CHexo}
\end{figure}

We note that giant planets of Jupiter-mass or greater can be detected in the outer regions of planetary systems using the direct imaging technique~\citep[e.g.,][]{vigan_sphere_2021}. Some debris disc systems have giants detected, such as the archetypal $\beta$ Pic system with its two planets~\citep{lagrange_constraining_2009, lagrange_evidence_2019}. For instance, it is interesting to note that some long-period planets detected, such as GJ 504 b or AF Lep b, have super-stellar metallicities, i.e. they appear to be enriched in CH$_4$. Those planets are more massive than Jupiter and we expect our scenario to only provide little carbon compared to their global masses, unless their primordial exo-KBs were significantly more massive than in our Solar System, or the atmosphere in which carbon can mix are relatively light. However, the effect on sub-Jupiter planets is expected to be much more significant and one can expect that late gas may contribute to the observed C/H in their atmospheres. Using our simulations, we can extrapolate the extra C/H ratio for an exoplanet as a function of its atmospheric mass (see Figure~\ref{fig:CHexo}).

For exoplanets, the question of the magnitude at which the different elements will exhibit an enrichment or not thus depends on the mass of the fluid envelope and that of the outer planetesimal belt as shown in Fig.~\ref{fig:CHexo} but also on the atmospheric temperature and entropy, something that is essentially linked to the planet's distance to the star. As we have seen the evaporation of CO gas from an outer belt should lead to a global increase of the deep atmospheric C and O abundances, but whether these are detectable depends on the atmospheric temperature itself. A strong difference between exoplanets and solar system giant planets is that for the latter we can probe their deep interior with measurements at microwave wavelengths that probe to tens or hundreds of bars~\citep{de_pater_possible_1991,bolton_jupiters_2017}, something impossible for exoplanets. For planets too hot for water clouds to form, both C and O should be enriched compared to other species such as N and S. For cooler exoplanets, we expect that the oxygen signature should be suppressed at photospheric temperatures lower than about 250\,K because of water condensation. For exoplanets with hydrogen atmospheres, the next major condensate is NH$_4$SH, which, mostly removes S from the upper atmosphere, due to its lower abundance for a solar-composition gas. This should occur around 150\,K. For still lower atmospheric temperatures, close to Jupiter's effective temperature (124\,K), NH$_3$ condenses and also removes N from the atmosphere. At this temperature and lower, interaction between water ice crystals and ammonia vapor leads to the formation of ``mushballs'' which can transport N even deeper~\citep{guillot_storms_2020}. When the photospheric temperature has decreased between Saturn and Uranus's values, i.e., between 95 and 60\,K, we expect that the transport by mushballs should decrease the abundance of N in the upper atmosphere below that of S~\citep{guillot_mushballs_2021}. This effectively implies that the NH$_4$SH condensation is not removing all of the S atoms, possibly leaving a possibility to observe it in the upper atmosphere. However, the condensation of H$_2$S, expected for planets with Uranus’ effective temperature and below (60\,K)~\citep{irwin_detection_2018} again removes S from the upper atmosphere, with H$_2$S which remains observable at depths for ice giants in the solar system using microwave wavelengths but would not be observable in an exoplanet. As can be seen, measuring a specific enrichment in carbon in exoplanets will require dedicated studies and a large sample.

As a conclusion, it appears that the simplest sample to study would be exoplanets with photospheric temperatures higher than 250 K. For these it is be expected that C, O, N and S would be observable in the upper atmosphere. In fact, observations of H$_2$S have already been made by the JWST for the super-Earth L 98-59 d~\citep{banerjee_atmospheric_2024}. Furthermore, HCN can also be observed with current observatories~\citep[e.g. the detection of HCN in HR 8799 c using VLTI/Gravity,][]{nasedkin_four---kind_2024}. With such observations of C, O, N and S available and growing over the next few years, one can expect to be able to quantify the amount of carbon accreted in the debris disc stage for a given planet. This is the way forward to test late gas accretion and decide on whether it could be the dominant mechanism that increases metallicity in planets.

The direct detection of less massive exoplanets at large distances from their stars will also grow thanks to the JWST sensitivity (using NIRCam and MIRI). Indeed, it may be possible to go down to Saturn-mass exoplanets or even lower for the closest planets to Earth reaching close to Neptune's mass~\citep{beichman_direct_2019}. When the first results arrive for those lower-mass planets, it will be interesting to see whether there is indeed a trend in metallicity and whether it may be linked to late gas accretion. However, we note that the limits in terms of detecting lines (e.g., water, methane, CO) for directly imaged planets with masses less than Jupiter are not yet clearly defined. Only future work or observations with the JWST will help us conclude~\citep[e.g.,][]{malin_simulated_2023}. The ELT is predicted to be able to image and characterize Saturn to Neptune-like exoplanets, notably with METIS, a first generation instrument~\citep{brandl_metis_2021}. It is also expected that second generation instruments such as PCS will be able to go down to exo-Earths to look for biosignatures and detecting Neptune-like planets may become routine~\citep{kasper_pcs_2021}.

On a longer timeframe, the Habitable Worlds Observatory~\citep[HWO,][]{national_academies_of_sciences_origins_2023} and the Large Interferometer for Exoplanets~\citep[LIFE,][]{quanz_atmospheric_2022} will be able to characterize those planets and further test the contribution of late gas as well as decipher any trend that may be present in terms of metallicity with the different characteristics of the systems.

\section{Conclusion}

It is now well known that gas (mostly CO) is released in exo-Kuiper belts around other stars. The gas originates from CO ices that sublimate over time from exo-Kuiper belt objects. It is likely that gas was also released in the young Kuiper belt in our Solar System. In this paper, we investigate the consequences of such gas release in the Solar System's youth on the atmospheres of Uranus and Neptune. Indeed, gas released at the Kuiper belt will spread inwards towards the giants, which will capture some of it. We have developed a numerical model capable of calculating the final masses of gas accreted onto Jupiter, Saturn, Uranus, and Neptune, assuming that the Solar System formed in a similar way to that proposed by the Nice model, but also for other scenarios that may explain the formation of the Solar System.

Assuming a massive primordial Kuiper belt similar to what is expected from state-of-the-art models, we find that up to $\sim$0.1 M$_\oplus$ of CO gas can be accreted onto each giant planet. The gas mixes within the upper atmospheres of giants and we find that it can totally change their primordial metallicities, in terms of observed C/H ratios. The effect is most impressive for Uranus and Neptune, as our model can explain [C/H] values of order 20 and 50, respectively as coming from capture of late gas. It means that our model may explain by itself the observed values of [C/H] on the ice giants. 

However, some of the observed [C/H] may originate from initial planet formation (e.g. accretion of planetesimals or pebbles). Using the S/H ratio as a tracer of this early accretion, we find that the extra [C/H] that may be needed to explain current observations is of order 20 for both Uranus and Neptune, which aligns very nicely with the results of our late gas capture model. We find that such a contribution can be provided with a primordial Kuiper belt more massive than 20-30 M$_\oplus$, similar to that proposed in the Nice model. We also find that constraints on the [C/H] in Jupiter and Saturn are compatible with such Nice-model scenarios. We conclude that late gas may play an important role in shaping the atmospheres of Uranus and Neptune as its contribution to the observed [C/H] may be dominant over early accretion during planet formation and is far from negligible even in worst-case scenarios. 

Furthermore, although this article has been devoted to the study of the implications of late gas in the solar system, the late gas mechanism we study is universal. Indeed, exo-Kuiper belts are present around more than 1/4 of stars and most of them could possess a gaseous component in their youth. Hence, exoplanets in planetary systems will suffer gas accretion (CO or C+O), which will inevitably alter their [C/H] ratios. We suggest some way forward to study the effect of late gas from currently accessible data (e.g. CH$_4$, HCN and H$_2$S observations in sub-Neptunes or super-earths) of sub-Jupiter planets close to their host star. Future observations of sub-Jupiter distant planets using direct imaging are challenging but may provide even more evidence of this late gas accretion, first the tip of the iceberg with the JWST, and then in a more refined fashion from the ground with the ELT, or with space missions like HWO or LIFE.

\begin{acknowledgements}
We thanks Philippe Thébault for interesting discussions on this paper.
\end{acknowledgements}

\bibliographystyle{aa}
\bibliography{Biblio}
\onecolumn
\begin{appendix}
\section{Extra figures and table}
\begin{table}[h]
    \caption{List of the Minimal and Maximal estimations for all LKB simulations of the atmospheric [C/H] ratio for Jupiter (J), Saturn, (S), Uranus (U), and Neptune (N) at 100 Myr of evolution.}
    \label{tab:LKB_results}
    \centering
    \begin{tabular}{c|c|c|c|c}
    \hline\hline 
        &  J & S & U & N \\
    \hline
    LKB-fid & $1.44 \ 10^{-6}$ -- $1.61 \ 10^{-6}$
    & $9.76 \ 10^{-6}$ -- $2.09 \ 10^{-5}$
    & $1.27 \ 10^{-4}$ -- $3.28 \ 10^{-4}$
    & $2.11 \ 10^{-4}$ -- $5.31 \ 10^{-4}$ \\
    LKB-2 & $2.03 \ 10^{-6}$ -- $2.27 \ 10^{-6}$
    & $1.36 \ 10^{-5}$ -- $2.89 \ 10^{-5}$
    & $1.68 \ 10^{-4}$ -- $4.36 \ 10^{-4}$
    & $2.42 \ 10^{-4}$ -- $6.79 \ 10^{-4}$ \\
    LKB-3 & $6.38 \ 10^{-7}$ -- $7.11 \ 10^{-7}$
    & $2.82 \ 10^{-6}$ -- $6.06 \ 10^{-6}$
    & $3.17 \ 10^{-5}$ -- $8.22 \ 10^{-5}$
    & $4.57 \ 10^{-5}$ -- $1.15 \ 10^{-4}$ \\
    LKB-4 & $1.16 \ 10^{-6}$ -- $1.29 \ 10^{-7}$
    & $1.15 \ 10^{-5}$ -- $2.46 \ 10^{-5}$
    & $1.77 \ 10^{-4}$ -- $4.59 \ 10^{-4}$
    & $3.40 \ 10^{-4}$ -- $8.57 \ 10^{-4}$ \\
    \hline
    \end{tabular}
\end{table}
\begin{figure}[htbp]
\begin{centering} 
\includegraphics[scale=0.57]{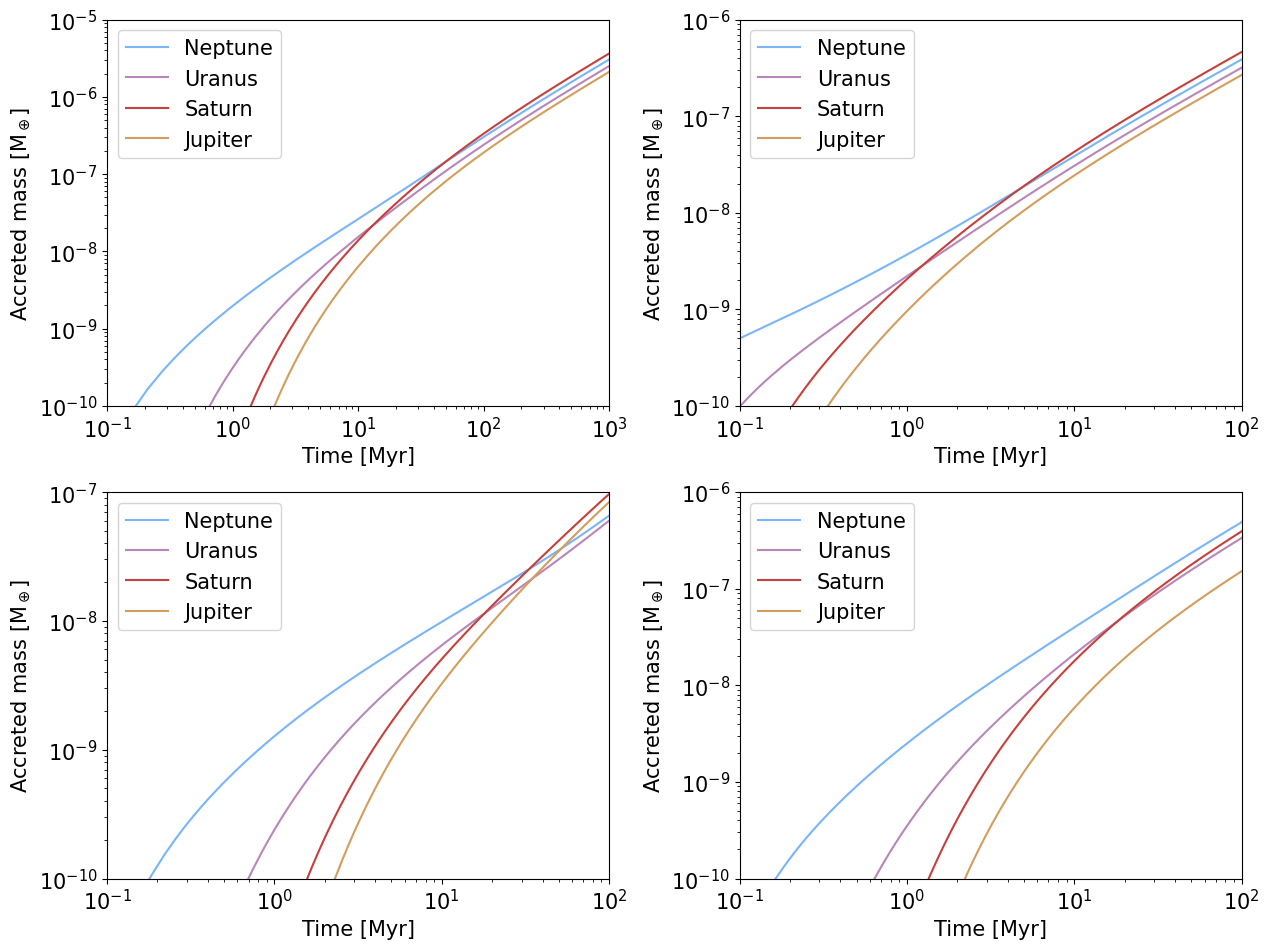}
\end{centering}
\caption{Neutral carbon accreted mass for each planet as a function of time for the simulations LKB-fid, 2, 3 and 4 (from left to right and top to bottom). The only differences between those simulations are the viscous parameter $\alpha$ and the accretion efficiency $f_{\rm accr}$ (see Table \ref{table:1}).}
\label{m_accr_LKB}
\end{figure}

\begin{figure}[tbh]
\begin{centering}
\includegraphics[scale=0.59]{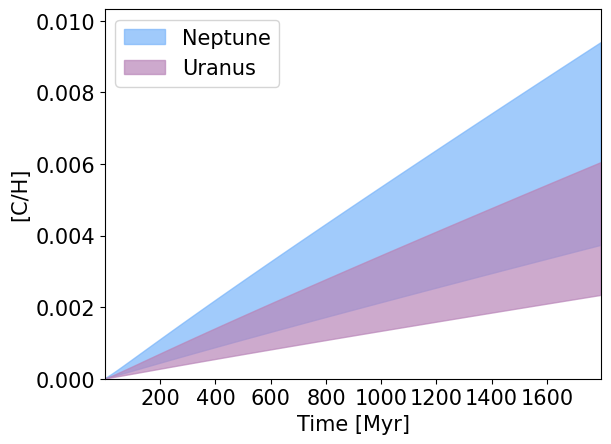}
\caption{[C/H] for the LKB-fid simulation as a function of time for the ice giants. The uncertainties represented by the filled areas are due to the uncertainties on the atmospheric masses of Neptune and Uranus.}
\end{centering}
\label{CH_LKB_fid}
\end{figure}
\begin{figure}[tbh]
\begin{centering}
\includegraphics[scale=0.60]{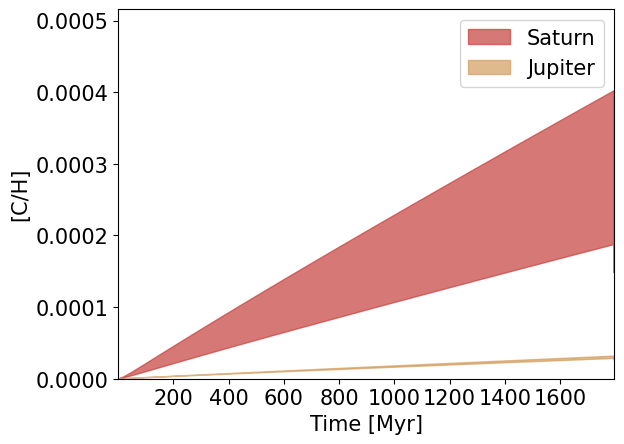}
\caption{C/H to protosolar ratio for the LKB-fid simulation as a function of time for the giant planets. The uncertainties represented by the filled areas are due to the uncertainties on the atmospheric masses of Jupiter and Saturn that may mix with carbon.}
\label{CH_LKB_fid_giants}
\end{centering}
\end{figure}
\end{appendix}
\end{document}